\documentclass[twocolumn]{aastex631}
\usepackage[utf8]{inputenc}
\usepackage[english]{babel}
\usepackage{amsmath}

\begin{document}

\title{Disk in the circumstellar envelope of carbon Mira V Cygni}

\correspondingauthor{Boris S. Safonov}
\email{safonov@sai.msu.ru}

\author[0000-0003-1713-3208]{Boris S. Safonov}
\affiliation{Sternberg Astronomical Institute Lomonosov Moscow State University \\
Universitetskii prospekt, 13, Moscow, Russian Federation, 119992}

\author[0000-0002-6321-0924]{Sergey G. Zheltoukhov}
\affiliation{Sternberg Astronomical Institute Lomonosov Moscow State University \\
Universitetskii prospekt, 13, Moscow, Russian Federation, 119992}

\author[0000-0002-4398-6258]{Andrey M. Tatarnikov}
\affiliation{Sternberg Astronomical Institute Lomonosov Moscow State University \\
Universitetskii prospekt, 13, Moscow, Russian Federation, 119992}

\author[0000-0003-0647-6133]{Ivan A. Strakhov}
\affiliation{Sternberg Astronomical Institute Lomonosov Moscow State University \\
Universitetskii prospekt, 13, Moscow, Russian Federation, 119992}

\author[0000-0002-2287-8151]{Victor I. Shenavrin}
\affiliation{Sternberg Astronomical Institute Lomonosov Moscow State University \\
Universitetskii prospekt, 13, Moscow, Russian Federation, 119992}

\begin{abstract}
AGB stars are the primary source of dust and complex molecules in the interstellar medium. The determination of outflow parameters is often hindered by the unknown geometry of the circumstellar environment, creating a demand for high-angular resolution observations. We use our NIR spectra and photometry of the carbon AGB star V~Cyg, along with literature data, to construct its SED over a wide range of wavelengths. The dust envelope responsible for the IR excess was also resolved in scattered polarized light at angular scales of 50--80~mas using differential speckle polarimetry. We present an interpretation of the thermal and scattered radiation of the dust using models of a spherical dusty outflow ($M_\mathrm{dust}=5.3\times10^{-7}$ M$_\odot$) and an inclined equatorial density enhancement, either in the form of a disk ($M_\mathrm{dust}=7.6\times10^{-3}$ M$_\oplus$) or a torus ($M_\mathrm{dust}=5.7\times10^{-3}$ M$_\oplus$), which material is concentrated at stellocentric distances less than 25~AU. The dust material consists of amorphous carbon and SiC, with 84\% of the dust being amorphous carbon. Dust particle radii range from 5 to 950~nm and follow a power law with an exponent of $-3.5$. Modeling of the envelope allowed us to improve the accuracy of stellar luminosity estimations: $21000 L_\odot$ and $8300 L_\odot$ at maximum and minimum brightness, respectively. The relation between the disk and the high water content in the envelope is also discussed.
\end{abstract}


\section{Introduction} \label{sec:intro}

Asymptotic giant branch (AGB) stars are the primary suppliers of gas and dust in the interstellar medium. At some stages of AGB a star can lose mass at rates up to $10^{-5}$~M$_\odot/yr$, and sometimes even $10^{-4}$~M$_\odot/yr$ \citep{Hofner2018}. It is widely accepted that the AGB is populated by stars with initial masses of $1-8$~M$_\odot$ (at solar metallicity, \citet{Hofner2018}). In the end of the AGB evolutionary stage, the star transitions to the so--called post--AGB stage, where it becomes a white dwarf surrounded by an atmosphere with a mass of $<0.01$~M$_\odot$. The difference in mass between the main--sequence star and the resulting white dwarf, amounting up to several solar masses, is released into interstellar medium as an extended gas--dust envelope.

The conditions in the cold, extended atmospheres of AGB stars are ideal for dust formation. The chemical composition of the dust depends on ratio of oxygen to carbon. For stars with initial masses in range of $1.7-4$~M$_\odot$, a process called the third dredge-up occurs \citep{Iben1983}, where newly formed carbon rises from the upper layers of the stellar core to the atmosphere, causing the C/O ratio to exceed 1. These stars, known as carbon stars, favour the formation of carbon--based dust particles, such as amorphous carbon or silicon carbide (SiC). Carbon stars can be easily identified by the presence of absorption lines of C$_2$, C$_2$H$_2$, HCN in their spectra. SiC can be detected by through its characteristic emission feature at 11.4~$\mu$m \citep{Hackwell1972}.

The high luminosities $L \sim 10^4$~L$_\odot$ and low effective temperatures $T_\mathrm{eff} \approx 3000$K of AGB stars result in their enormous sizes, often exceeding hundreds of solar radii. This, in turn, leads to a low gravitational force at the photosphere (log~$g \sim 0-1$), facilitating the formation of stellar wind. Newly formed dust particles in the upper atmosphere of the star are accelerated by star radiation pressure and drag gas along with them. The cold, extended atmospheres of AGB stars are an significant source of various molecules found in the interstellar medium.

V~Cyg is one of stars populating AGB. It is a bright carbon Mira variable star of spectral type C7,4, with an effective temperature of $T_\mathrm{eff}=2510$~K and a high luminosity $L_{\odot}\approx15000$L$_\odot$ \citep{Cohen1979}. Different estimates of V Cyg luminosity and temperature, derived from various data and assumptions, can be found in the literature (e.g., \citealt{Bergeat2001,vanBelle1997,Groenewegen1998}). Temperature estimates range from 1800 to 2500~K, and luminosity  estimates vary from 7000 to 25000~L$_\odot$. The mass of V~Cyg is near the upper end of the range typical for carbon stars, around $2 - 4$~M$_\odot$. The parallax of V~Cyg, according to Gaia DR3, is $1.8331 \pm 0.1454$~mas \citep{gaiadr3}, from which we adopt a distance of $565$~pc. It is noteworthy that V~Cyg has a relatively high Gaia Renormalized Unit Weight Error of 6.187, which may be attributed to high--amplitude variability or binarity of the object.

According to GCVS \citep{Samus2017}, V~Cyg pulsates with a period of $P=421.27^d$, with its magnitude varying between $V\sim8^m$ and $13^m$. Using the period--luminosity relationship for Mira variables \citep{Whitelock2008}, we obtain an absolute magnitude of $M_K\approx-8^m$. Applying a bolometric correction $BC_K\sim3$ \citep{Kerschbaum2010} results in a luminosity estimate of $\sim8000$~L$_\odot$.

A circumstellar dust envelope around V~Cyg was detected in the mid-IR by \citet{Forrest1975}, with its SED showing a prominent emission feature at $\lambda=11.3$~$\mu$m, attributed to SiC dust particles \citep{Treffers1974}. The dust envelope  parameters of V~Cyg have been estimated by studies such as \citet{Groenewegen1998,Etmanski2020}. These studies assumed a black body model for the stellar spectrum, though a model using a more realistic spectrum (e.g. from \citep{Aringer2009})  has yet to be developed. Existing models of the V~Cyg envelope also do not account for ISO spectra in the $2.4-45~\mu$m range obtained in 1997. 

V~Cyg exhibits significant, variable intrinsic polarization, reaching up to $4.3\%$ \citep{Melikian1996} (measured in an unspecified band, likely close to the visual range). In evolved stars, polarized radiation arises from scattering in the dust envelope. For a centrally symmetrical envelope, the total polarization averages to zero due to the symmetrical azimuthal polarization pattern. However, the non--zero polarization observed in V~Cyg indicates asymmetry in the envelope shape. For instance, scattering on an asymmetric structure, such as a disk or dust cloud on one side of the system, would produce non--zero integrated polarization \citep{Whitney1992,Shrestha2021}. However, with only two values --- the degree and angle of polarization --- it is impossible to distinguish between these scenarios or draw quantitative conclusions. Furthermore, uncertainty in envelope geometry can introduce biases in mass--loss estimates from the SED \citep{Wiegert2020}. At the same time, resolved scattered polarized radiation can offer crucial quantitative constraints on the object model, it was done before for evolved stars by \citet{Norris2012,Scicluna2015,Safonov2019b,Fedoteva2020}. 

In this work, we combine our observations of V~Cyg with data from the literature to construct a model of the circumstellar environment and to estimate the object's luminosity with improved precision. Section~\ref{subs:SED} describes observations of the object using IR photometry and IR spectroscopy. Differential Speckle Polarimetry (DSP) was employed to study scattered radiation with an angular resolution of 50--80~mas in visual wavelengths, as detailed in subsection~\ref{subs:DPVobs}. In section~\ref{sec:SED} we examine the object's variability and construct the SED. High--angular resolution image of the object in polarized light is reconstructed from DSP data in section~\ref{subs:img}. We interpret the array of observational data using Monte Carlo Radiation Transfer modeling of the circumstellar envelope in section~\ref{sec:model}. The results of modeling are discussed in section~\ref{sec:discussion}. Section~\ref{sec:conclusion} presents conclusions of the study.

\section{OBSERVATIONS}
\label{sec:obs}

\subsection{Photometry and spectroscopy}
\label{subs:SED}

\begin{figure}[t!]
  \center
  \plotone{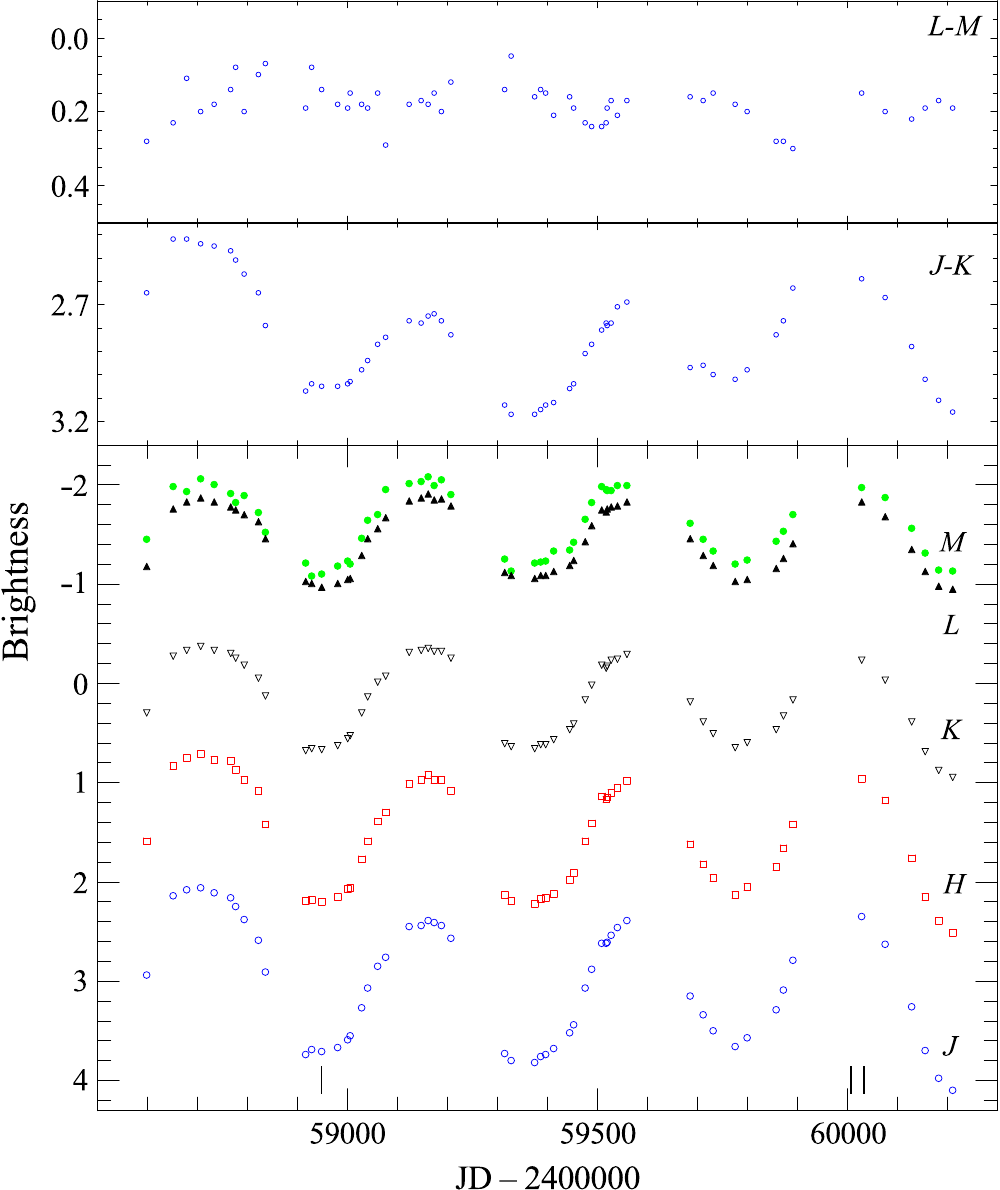}
  \caption{Lower part: $JHKLM$ light curves. Upper part: color indices $J-K$, $L-M$. Short vertical lines above the horizontal axis indicate the moments of spectral observations.}
  \label{fig:jhklm}
\end{figure}

$JHKLM$ photometry was obtained using the single channel InSb--photometer mounted on the 1.25-m telescope ZTE of the Crimean Station of the Sternberg Astronomical Institute (SAI) of Lomonosov MSU. The results of these observations are presented in Figure~\ref{fig:jhklm} and Table~\ref{table:jhklm_tab}. The star BS\,7924, with magnitudes $J=1.00$, $H=0.91$, $K=0.89$, $L=0.78$, $M=0.87$, was used as a photometric standard. The precision of estimates in all bands is better than $0.02^m$.

\begin{deluxetable}{cccccc}[b]
\tablecaption{$JHKLM$ photometry of V~Cyg (full version is available in electronic form).
\label{table:jhklm_tab}}
\tablewidth{0pt}
\tablehead{\colhead{JD 2400000+} & \colhead{$J$} & \colhead{$H$} & \colhead{$K$} & \colhead{$L$} & \colhead{$M$}}
\startdata
  58598.6 & 2.94 & 1.59 &  0.29 & -1.17 & -1.45 \\
  58651.5 & 2.14 & 0.83 & -0.28 & -1.75 & -1.98 \\
  58678.5 & 2.08 & 0.75 & -0.34 & -1.82 & -1.93 \\
\enddata
\end{deluxetable}

NIR (1-2.5~$\mu$m) spectra of V~Cyg were obtained using the ASTRONIRCAM instrument \citep{Nadjip2017}, mounted on the Nasmyth--1 platform of the 2.5-m telescope at the Caucasian Mountain Observatory (CMO, \citet{Shatsky2020}) of SAI MSU. A detailed description of the spectral mode of ASTRONIRCAM is provided by \citet{Zheltoukhov2020}. Observations were conducted in cross--dispersion mode using $0.9^{\prime\prime}$ wide slit. A0V stars located at similar altitudes as the object at the time of observations were used as telluric standards. Absolute flux calibration was not performed. The typical effective spectral resolution is $\approx1100$. The moments of spectral observations are listed in Table~\ref{tab:V_Cyg_sp_CMO} and marked in Figure~\ref{fig:jhklm}.

\begin{deluxetable}{ccccccc}[t!]
\caption{Observational log for NIR spectra obtained using ASTRONIRCAM.
\label{tab:V_Cyg_sp_CMO}}
\tablewidth{0pt}
\tablehead{\colhead{UT} & \colhead{MJD} & \colhead{phase} & \colhead{$t_\mathrm{tot}$, s} & \colhead{$t_\mathrm{tot}$, s} & \colhead{SNR} & \colhead{SNR} \\
& & & \colhead{$YJ$} & \colhead{$HK$} & \colhead{$YJ$} & \colhead{$HK$}
}
\startdata
2020-04-08 & 58947.0 & 0.53 & 195 & 88 & 136 & 95 \\
2023-03-03 & 60006.1 & 0.03 & 195 & 51 & 58 & 60 \\
2023-03-29 & 60032.1 & 0.09 & 293 & 88 & 80 & 92 \\
\enddata
\end{deluxetable}

\subsection{Differential Speckle Polarimetry}
\label{subs:DPVobs}

We observed V~Cyg using the SPeckle Polarimeter \citep{Safonov2017} on September 21st, 2020 ($JD=2459114$). The instrument was mounted on the 2.5-m telescope of CMO SAI MSU. The epoch of observation is $\approx50^d$ before the maximum brightness in the $J$ band. The observations were conducted at three bands centered at 550, 625, and 880~nm, with half--widths of 50, 50, and 70~nm, respectively. Seeing conditions, as measured by MASS--DIMM, ranged between $1.1-1.2^{\prime\prime}$ \citep{Kornilov2016}. The angular scale of the speckle polarimeter camera was $20.6$~mas/pix, with the scale and orientation calibrated using observations of known binary stars. Each individual frame had an exposure time of 30~ms. Additional details of the observations are provided in Table~\ref{table:obslog}.

\begin{deluxetable}{ccccccc}[t]
\tablecaption{Observational log for DSP.
\label{table:obslog}}
\tablewidth{0pt}
\tablehead{\colhead{UT} & \colhead{$\lambda_c$} & \colhead{FOV} & \colhead{FOV} & \colhead{$G_\mathrm{EM}$} & \colhead{$t_\mathrm{tot}$} & \colhead{$n_e$} \\
 & \colhead{nm} & pix & \colhead{$^{\prime\prime}$} & & \colhead{s} & \\
\colhead{(1)} & \colhead{(2)} & \colhead{(3)} & \colhead{(4)} & \colhead{(5)} & \colhead{(6)} & \colhead{(7)} 
}
\startdata
18:30:56 & 880 & 134 & 2.76 & 78 & 206 & $4.2\times10^5$ \\ 
18:34:46 & 625 & 148 & 3.05 & 202 & 206 & $1.8\times10^5$ \\ 
18:38:36 & 550 & 150 & 3.09 & 500 & 207 & $4.5\times10^4$ \\ 
\enddata
\tablecomments{All observations were conducted on September 21th, 2020. Columns: 1 --- central moment of series, 2 --- central wavelength of employed filter, 3 --- field of view in pixels, 4 --- field of view in arcseconds, 5 --- electron multiplication, 6 --- total integration time, 7 --- average number of photoelectrons in single frame.
}
\end{deluxetable}

\begin{figure*}[ht]
  \center
  \plotone{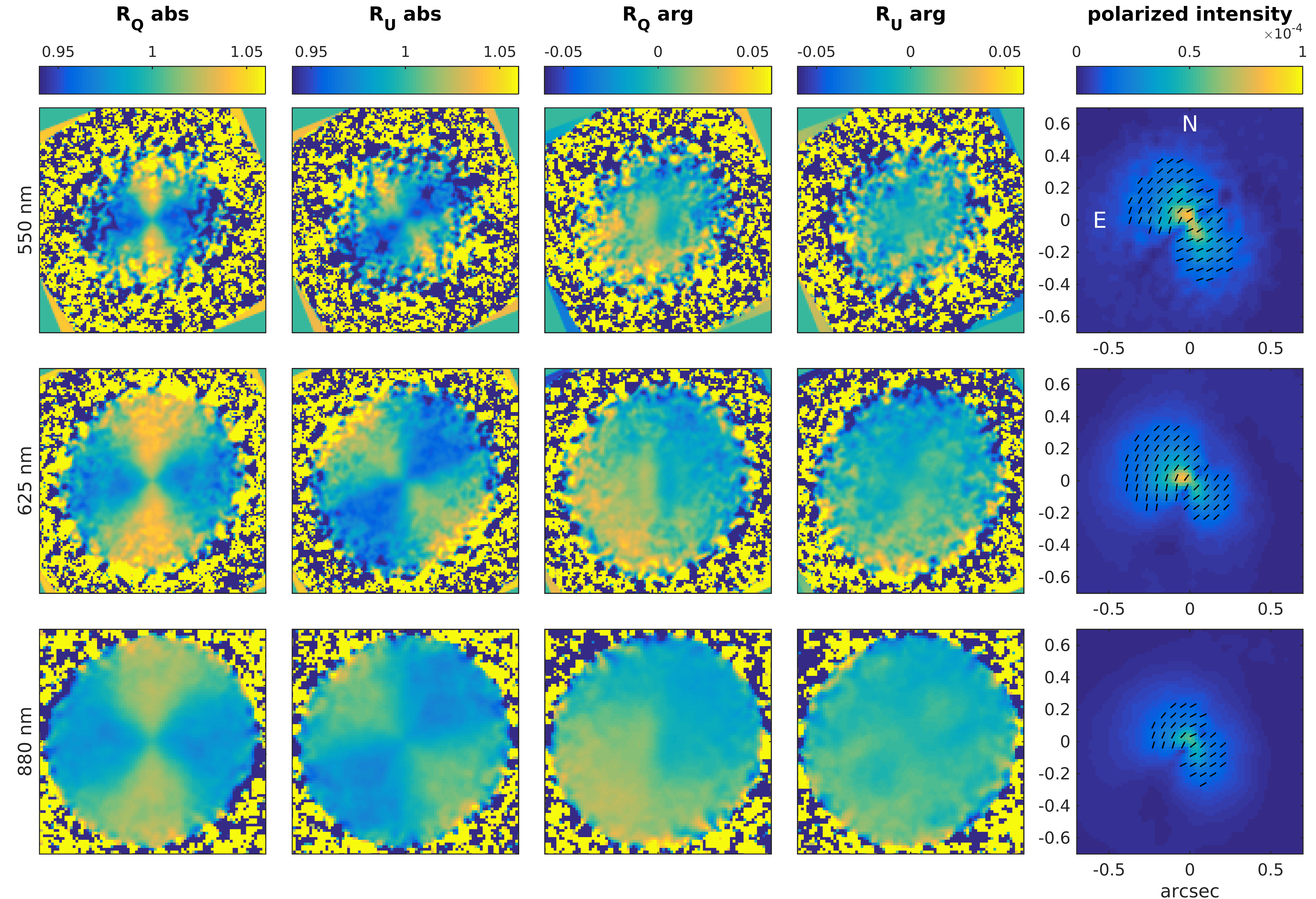}
  \caption{Differential polarimetric visibility of V~Cyg. The rows correspond to observations in the 550, 625, and 880~nm bands, from top to bottom. The first four columns display the following DPV components: $|R_Q|$, $|R_U|$, $\mathrm{arg}R_Q$, $\mathrm{arg}R_U$. Spatial frequency is shown along the axes, with the displayed domain having a size of $2D/\lambda$, where $D$ is the telescope diameter and $\lambda$ is the wavelength. The fifth column presents the image in polarized intensity, reconstructed as described in Section~\ref{subs:img}. The units of intensity are flux per $20\times20$ pixel, relative to the total stellar flux. Dashes indicate the orientation of the polarization plane. In all panels, North is up, East is to the left.}
  \label{fig:DSPobs}
\end{figure*}

The principal observable in Differential Speckle Polarimetry (DSP) is the differential polarization visibility (DPV), defined as the ratio of visibility of the object in orthogonal polarizations \citep{Norris2012}:
\begin{equation}
\mathcal{R}_Q(\boldsymbol{f}) = \frac{\widetilde{I}(\boldsymbol{f})+\widetilde{Q}(\boldsymbol{f})}{\widetilde{I}(\boldsymbol{f})-\widetilde{Q}(\boldsymbol{f})},\,\,\,\mathcal{R}_U(\boldsymbol{f}) = \frac{\widetilde{I}(\boldsymbol{f})+\widetilde{U}(\boldsymbol{f})}{\widetilde{I}(\boldsymbol{f})-\widetilde{U}(\boldsymbol{f})}.
\label{eq:Rdef}
\end{equation}
Here, the tilde symbol denotes the Fourier transform, $\boldsymbol{f}$ is the two--dimensional spatial frequency vector, and $I, Q, U$ are the distributions of the Stokes parameters of the object. The DPV is estimated from a series of short--exposure images, obtained simultaneously in two orthogonal polarizations  without correction for atmospheric distortions, using the full telescope aperture \citep{Safonov2019}. Pairs of simultaneous short--exposure images are used to estimate their subtle differences caused by polarization, while their speckle patterns induced by the atmospheric turbulence are identical. Averaging of difference--dependent quantities in the Fourier space allows us to extract the polarized-light image structure close to diffraction-limited resolution.

The speckle polarimeter is equipped with a half--wave plate (HWP) that acts as a modulator. During the series acquisition, the HWP rotates continuously, with its angular speed chosen so that the detector records exactly 40~frames per HWP revolution. An electron Multiplying CCD Andor iXon 897 is used as the detector.

\begin{figure*}[t!]
  \center
  \includegraphics[width=0.85\linewidth]{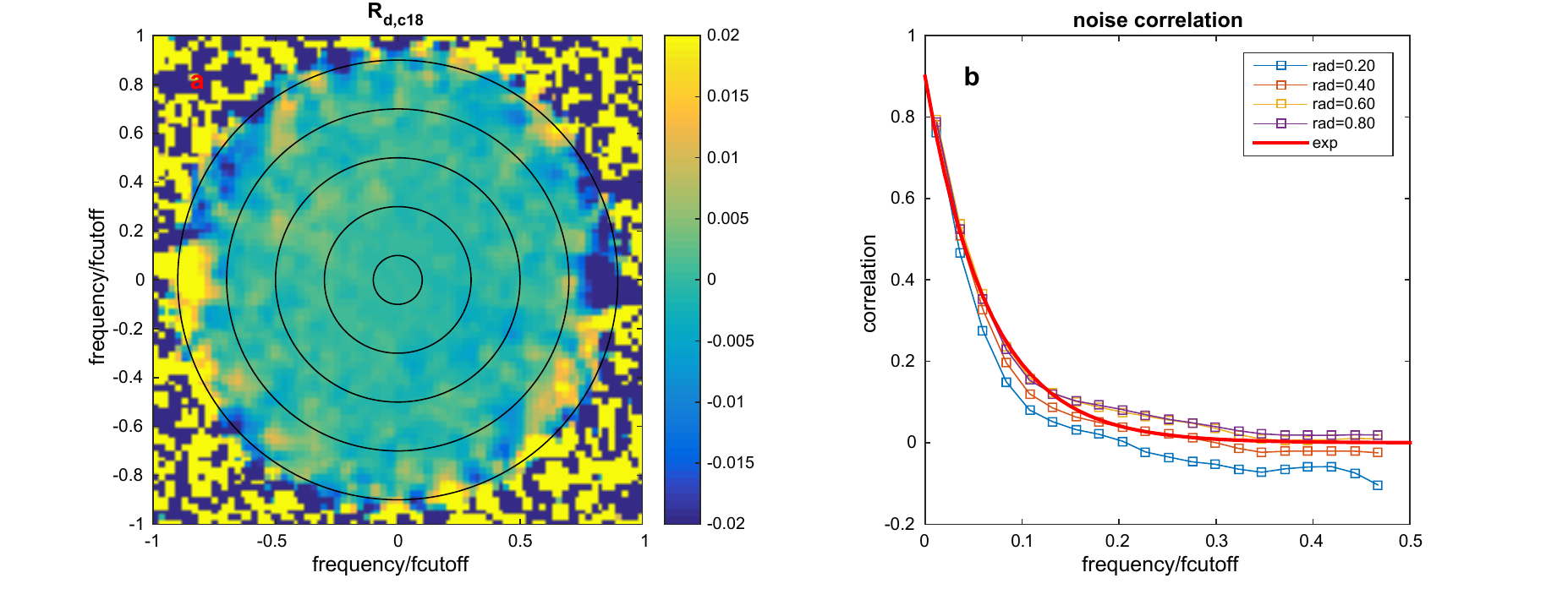}
  \includegraphics[width=0.85\linewidth]{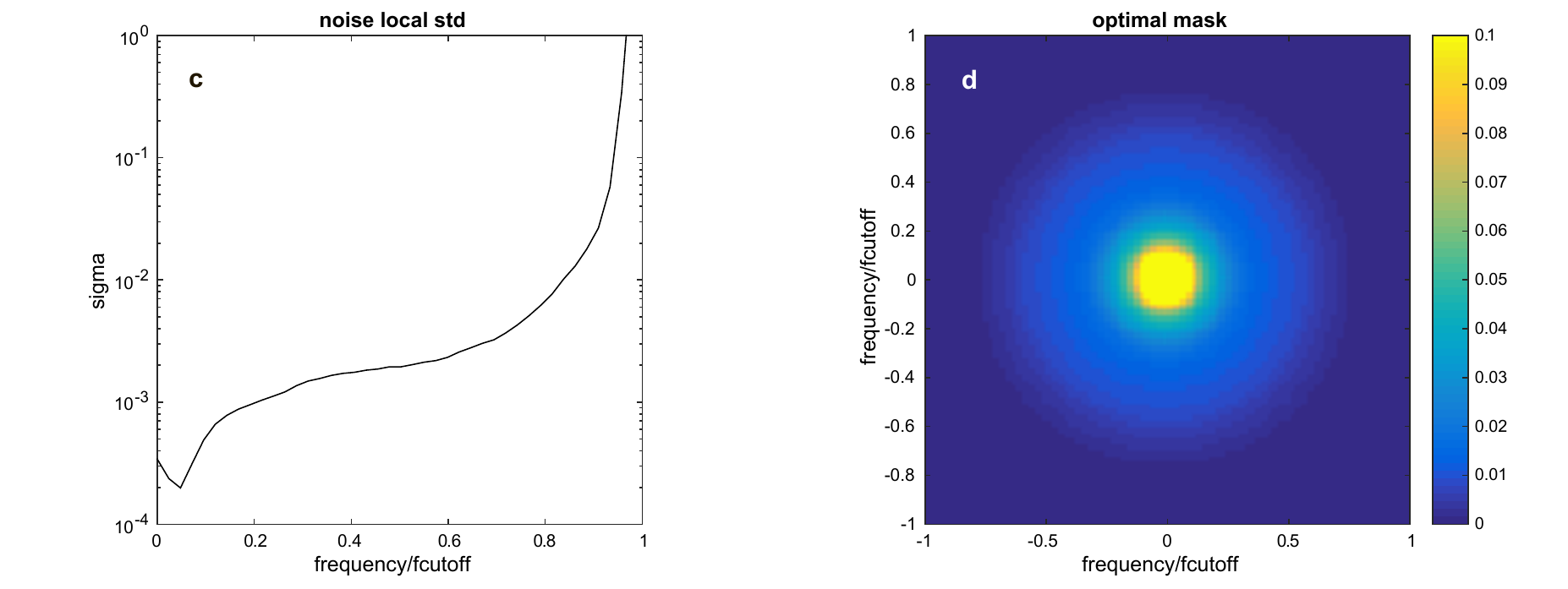}
  \caption{Noise characteristics of DPV $\mathcal{R}_Q$, $\mathcal{R}_U$ for observations at 880~nm. {\bf a.} $\mathcal{R}$ demodulated using the factor $\cos18\gamma$. Black lines indicate the ring zones used later in Section~\ref{subs:DPVnoise} for the noise correlation analysis. {\bf b.} The two--point correlation of the noise as a function of the distance between the considered points. Different lines correspond to different ring zones (as described above). The red line shows the function $\exp(-f_r/f_l)$ at $f_l=0.065f_c$. {\bf c.} Local standard deviation of noise as a function of spatial frequency.  {\bf d.} Spectral filter used for image reconstruction.}
  \label{fig:DSPnoise}
\end{figure*}

The measurements results are presented in the first four columns of Figure~\ref{fig:DSPobs}. Instrumental polarization effects were corrected using the method described by \citep{Safonov2019}. Given the small interstellar extinction for this object (see Section~\ref{subs:SED}), the potential influence of interstellar polarization can be neglected. In all three bands, we detected significant deviation in DPV amplitude from unity and DPV phase from zero. The former indicates that the polarized flux is reliably resolved, while the latter confirms that it lacks central symmetry.

The extraction of DPV from a series of short--exposure images is performed via demodulation. While the full details of the algorithm are described in \citep{Safonov2019}, we highlight one aspect critical to this study. Denoting the position angle of the HWP as $\gamma$, the demodulation process involves averaging the raw DPV signal multiplied by the factors  $\cos N\gamma$, $\sin N\gamma$. The harmonic $N=4$ carries useful signal, with $\cos4\gamma$ corresponding to $\mathcal{R}_Q$, $\sin4\gamma$ corresponding to $\mathcal{R}_U$. Harmonics with $4<N<20$ are also important, as they contain noise realizations. An example of a noise harmonic (N=18, $\cos$--demodulation) for observations at $\lambda=880$~nm is shown in Figure~\ref{fig:DSPnoise}, panel (a). To estimate the noise of measurements $\sigma^2$ at each point of Fourier space, we computed the variance across harmonics $4<N<20$.

\section{Spectral Energy Distribution}
\label{sec:SED}

\subsection{Photometric and spectral variability}

The $JHKLM$ light curves for V~Cyg are presented in Fig.~\ref{fig:jhklm}. They clearly exhibit periodic flux fluctuations, with amplitude decreasing at longer wavelengths. The amplitude changes from period to period by $0.2-0.5^m$, with the effect being most prominent at the $J$ band. This behavior is typical for similar objects \citep{Fedoteva2020, Tatarnikov2024}. Additionally, the visual light curve, available on the AAVSO website\footnote{https://www.aavso.org/LCGv2/}, reveals variations in both the average magnitude and the amplitude of fluctuations on a characteristic timescale of $\sim12$~yr. The color indices computed for the $JHK$ bands are larger during minimum brightness.

\begin{figure}[ht]
  \center
  \includegraphics[width=0.85\linewidth]{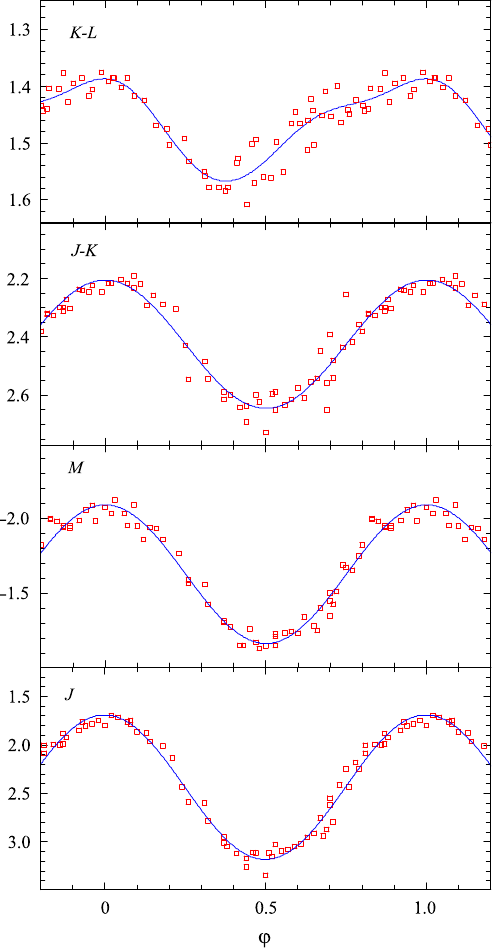}
  \caption{Phase curves of $J$, $M$ magnitudes and $J-K$, $K-L$ color indices. Blue lines show approximation by Fourier series containing terms of order of $\leq3$.}
  \label{fig:phot_phase}
\end{figure}

To study the periodicity of the brightness variations, the data were first corrected using a third degree polynomial to remove secular trends. The residuals were then decomposed into Fourier series up to the third harmonics. No significant phase shift between the light curves in different bands was detected (Fig.~\ref{fig:phot_phase}). The ephemeris for maximum brightness are as follows: $\text{JD}_{max}=2458721_{\pm12}+423.8_{\pm 4}\cdot E$. Color indices variations align with these ephemeris, with phase $\varphi=0$ corresponding to the minimum of a color index. For shorter-wavelength bands, the color phase curves exhibit a symmetrical shape with a maximum at $\varphi=0.5$. In contrast, index $K-L$ phase curve shows significant asymmetry, with its maximum occuring at $\varphi=0.38$. The amplitudes of magnitude after detrending are $\Delta J=1.48^m$, $\Delta H=1.35^m$, $\Delta K=1.05^m$, $\Delta L=0.92^m$, $\Delta M=0.94^m$. These values are typical for Mira variables, showing a decrease in amplitude from $J$ to $K$ and relative constancy for $L$ and $M$ bands, consistent with findings in \citet{Fedoteva2020} and references therein. 

\begin{figure}[ht]
  \center
  \includegraphics[width=0.85\linewidth]{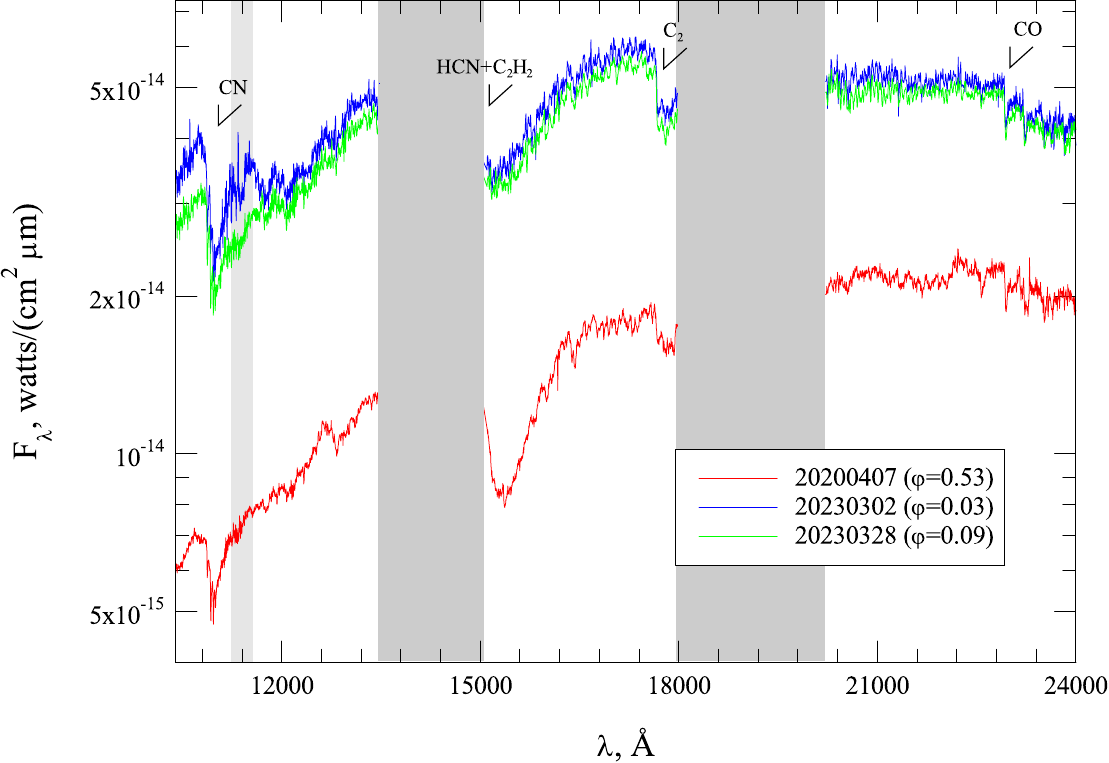}
  \caption{NIR spectra of V Cyg at maximum and minimum of brightness. The regions of strong telluric absorption bands are shaded in grey. The light grey band at a $\lambda \approx 1.14$ $\mu$m marks the region, in which signal recovery is possible.}
  \label{fig:spectra}
\end{figure}

The spectra of V~Cyg, shown in Figure~\ref{fig:spectra}, were obtained at phases near the maximum and minimum of brightness. These spectra prominently display molecular bands such as C$_2$, CO, CN, which are characteristic of carbon stars. A notable phase-dependent behavior is observed in the depth of the C$_2$ and HCN+C$_2$H$_2$. Near the brightness minimum, the C$_2$ band diminishes in prominence, while the HCN+C$_2$H$_2$ band, on the contrary, becomes stronger. In contrast, the CO bands ($\lambda=2.29$ $\mu$m) remain clearly visible throughout and show no significant phase dependence in their depth. This constancy likely indicates that the CO bands are formed within the extended dust envelope rather than in the stellar atmosphere itself \citep{Tatarnikov2024}.

\subsection{SED compilation}

The notably large color index $J-K \sim 3$ observed for V~Cyg suggests either a large interstellar absorption in the object direction or the presence of a circumstellar dust envelope. Using the distance to the star from GAIA DR3 \citep{gaiadr3} and the interstellar absorption map of \citet{dustmap}, we expect interstellar color excess $E(g-r)=0.24^m$. Following the standard  interstellar reddening law, this corresponds to $E(J-K) \approx 0.13^m$. This value is far too small to account for the observed values of $J-K$. Thus, the large $J-K$ color index is likely dominated by the effects of a dense circumstellar dust envelope, which contributes to reddening and alters the spectral energy distribution of the star.

\begin{figure}[t]
  \center
  \includegraphics[width=1.0\linewidth]{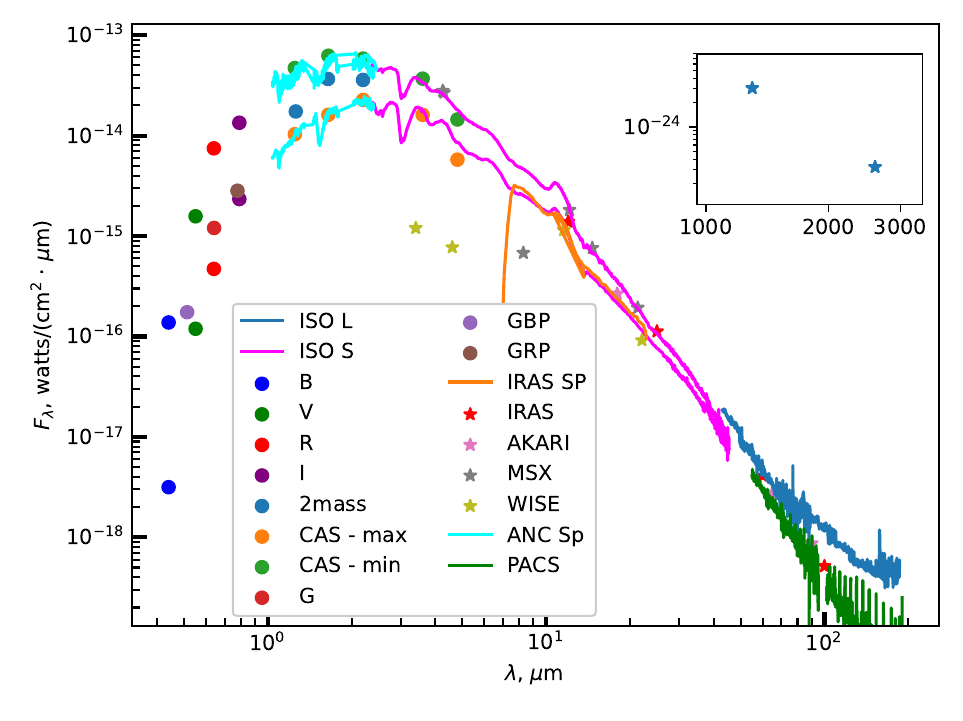}
  \caption{SED of V Cyg at the brightness maximum and minimum. The inset shows two flux measurements made with IRAM \citep{CastroCarrizo2010}.}
  \label{fig:sed_all}
\end{figure}

To construct the spectral energy distribution of V~Cyg, we utilized data from various sources: spectra from ISO \citep{isohpdp} and IRAS \citep{iras}, photometric observations from the 2MASS survey \citep{2mass}, IRAS, AKARI \citep{akari}, WISE \citep{WISE}, MSX \citep{msx}, IRAM \citep{CastroCarrizo2010}, GAIA \citep{gaia1} and \citep{gaiadr3}, the Tycho-2 catalogue \citep{tycho2}, and the AAVSO database. Additionally, we incorporated our own observations obtained at CAS and CMO of the Moscow University. The Fig.~\ref{fig:sed_all} shows the compiled data for both the minimum and maximum brightness phases. 

For the analysis we used data obtained near the maximum phase, correcting for interstellar absorption using the reddening law from \citet{ccm89} with $A_V=0.74^m$. By integrating the spectrum over a wide spectral range, we derived the bolometric flux $F_\mathrm{bol}^\mathrm{max}=2.3\cdot10^{-9}$ w/m$^2$, leading to an estimated luminosity of $\approx21000$~L$_\odot$. Using the angular diameter of V~Cyg ($\theta \approx 13$ mas)  reported by \citet{angsize}, we estimated the effective temperature $T_\mathrm{eff} \approx 2600$~K.

\section{Reconstruction of image in polarized flux}
\label{subs:img}

In this section, we use DPV measurements to reconstruct images of the circumstellar envelope of V~Cyg in polarized intensity. We model the object as a combination of point--like source (the star) and an extended envelope. Since the direct stellar radiation is unpolarized, the Fourier transforms of Stokes parameters distributions can be expressed as:
\begin{equation}
\widetilde{I}(\boldsymbol{f}) = \widetilde{I}_\star + \widetilde{I}_\mathrm{e}(\boldsymbol{f}),
\label{eq:objectI}
\end{equation}
\begin{equation}
\widetilde{Q}(\boldsymbol{f}) = \widetilde{Q}_\mathrm{e}(\boldsymbol{f}),
\label{eq:objectQ}
\end{equation}
\begin{equation}
\widetilde{U}(\boldsymbol{f}) = \widetilde{U}_\mathrm{e}(\boldsymbol{f}).
\label{eq:objectU}
\end{equation}
Here $\widetilde{I}_\star$ is the Fourier transform of the star, which is constant across all spatial frequencies and equals the stellar flux $I_\star$. The terms $\widetilde{I}_\mathrm{e}(\boldsymbol{f})$, $\widetilde{Q}_\mathrm{e}(\boldsymbol{f})$, and $\widetilde{U}_\mathrm{e}(\boldsymbol{f})$ are the Fourier transforms of the envelope intensity distributions in Stokes $I$, $Q$, $U$, respectively.

Assuming that the envelope is significantly fainter than the star $\widetilde{Q}_\mathrm{e}\ll\widetilde{I}_\star$, $\widetilde{U}_\mathrm{e}\ll\widetilde{I}_\star$, we can substitute the expressions above into definitions of DPV and retain only first-order terms. This yields the following for the normalized Fourier transforms of the envelope:
\begin{equation}
\widetilde{Q}_\mathrm{e}^{\prime}(\boldsymbol{f}) = 0.5\bigl(\mathcal{R}_Q(\boldsymbol{f})-1\bigr),
\end{equation}
\begin{equation}
\widetilde{U}_\mathrm{e}^{\prime}(\boldsymbol{f}) = 0.5\bigl(\mathcal{R}_U(\boldsymbol{f})-1\bigr).
\end{equation}
Here, the quantities $\widetilde{Q}_\mathrm{e}^{\prime}(\boldsymbol{f})$ $\widetilde{U}_\mathrm{e}^{\prime}(\boldsymbol{f})$  are normalized by the total stellar flux of the star $\widetilde{I}_\star$, such that $\widetilde{Q}_\mathrm{e}^{\prime}(\boldsymbol{f})=(\widetilde{Q}_\mathrm{e}(\boldsymbol{f})/I_\star)$, $\widetilde{U}_\mathrm{e}^{\prime}(\boldsymbol{f})=(\widetilde{U}_\mathrm{e}(\boldsymbol{f})/I_\star)$.

In principle, the images of the envelope in Stokes $Q$ and $U$ can be reconstructed by applying the inverse Fourier transform to $\widetilde{Q}_\mathrm{e}^{\prime}(\boldsymbol{f})$ and $\widetilde{U}_\mathrm{e}^{\prime}(\boldsymbol{f})$. However, as shown in Figure~\ref{fig:DSPobs} and by the noise analysis conducted in Section~\ref{subs:DPVnoise}, $\mathcal{R}_Q$ and $\mathcal{R}_U$ and, consequently, $\widetilde{Q}_\mathrm{e}^{\prime}(\boldsymbol{f})$ and $\widetilde{U}_\mathrm{e}^{\prime}(\boldsymbol{f})$ have an acceptable signal--to--noise ratio only at frequencies below 0.6-0.9$f_c$.

To mitigate the noise at higher frequencies, we applied a spectral filter $G_\mathrm{opt}(\boldsymbol{f})G(\boldsymbol{f})$ to $\widetilde{Q}_\mathrm{e}^{\prime}(\boldsymbol{f})$ and $\widetilde{U}_\mathrm{e}^{\prime}(\boldsymbol{f})$. Here  $G_\mathrm{opt}(\boldsymbol{f})$ is the optimal filter which is computed from the uncertainty of $\mathcal{R}$ (as described in the end of Section~\ref{subs:DPVobs}):
\begin{equation}
    G_\mathrm{opt}(\boldsymbol{f}) = 1/\sigma^2(\boldsymbol{f}),
\end{equation}
where $\sigma^2(\boldsymbol{f})$ is the variance of $\mathcal{R}$ at each spatial frequency. Additional filter $G(\boldsymbol{f})$ is the diffraction-limited optical transfer function of the aperture, defined as:
\begin{equation}
G(\boldsymbol{f})=(2/\pi)\Bigl[\mathrm{arccos}z-z\sqrt{1-z^2} \Bigr],
\end{equation}
where $z=2|\boldsymbol{f}|\lambda/D^{\prime}$. The resulting combined spectral filter for the observations at $880$~nm is shown in the lower-right panel of Figure~\ref{fig:DSPnoise}.

We apply the inverse Fourier transform to the quantities $G_\mathrm{opt}(f)G(f)\widetilde{Q}_\mathrm{e}^{\prime}(\boldsymbol{f})$ and $G_\mathrm{opt}(f)G(f)\widetilde{U}_\mathrm{e}^{\prime}(\boldsymbol{f})$. As long as $\widetilde{Q}_\mathrm{e}^{\prime}(\boldsymbol{f})$ and $\widetilde{U}_\mathrm{e}^{\prime}(\boldsymbol{f})$ contain useful information up to frequencies $0.6-0.9f_c\equiv0.6-0.9D/\lambda$, the resulting images $Q^{\prime}(\boldsymbol{\alpha})$ and $U^{\prime}(\boldsymbol{\alpha})$ achieve an angular resolution $1.1-1.7\lambda/D$, which is close to the diffraction limit. Here, $\alpha$ is a two-dimensional angular coordinate, with its origin set at the point--like unpolarized star.

The $Q^{\prime}(\boldsymbol{\alpha})$ and $U^{\prime}(\boldsymbol{\alpha})$ maps were used to compute the polarized intensity $I_p(\boldsymbol{\alpha})$ and the angle of polarization $\chi(\boldsymbol{\alpha})$:
\begin{equation}
I_p(\boldsymbol{\alpha}) = \sqrt{Q^{\prime2}(\boldsymbol{\alpha})+U^{\prime2}(\boldsymbol{\alpha})},
\end{equation}
\begin{equation}
\chi(\boldsymbol{\alpha}) = \frac{1}{2}\mathrm{arctg}\bigl(U^{\prime}(\boldsymbol{\alpha})/Q^{\prime}(\boldsymbol{\alpha})\bigr).
\end{equation}
Recall that polarized intensity equals the total intensity multiplied by the degree of polarization. The described method of polarized image reconstruction has been applied previously to the protoplanetary disk of CQ~Tau \citep{Safonov2022} and the circumstellar environment of Betelgeuse \citep{Safonov2020}. In both cases resulting images were consistent with observations conducted using larger telescopes.

The $I_p(\boldsymbol{\alpha})$ and $\chi(\boldsymbol{\alpha})$ maps for V~Cyg are presented in the fifth column of Figure~\ref{fig:DSPobs}. A polarized envelope is visible at stellocentric angular distances of $0.05-0.5^{\prime\prime}$, with brightness gradually decreasing outward. It is likely that the envelope extends further, however, the DSP method lacks the sensitivity to detect it at distances greater than $0.5^{\prime\prime}$. The envelope exhibits an azimuthal polarization pattern, which unambiguously indicates that it is a reflection nebula. Two lobes of increased brightness are apparent --- one to the northeast and another to the southwest. The appearance of the envelope remains consistent across different bands. The reconstructed images will primarily guide the selection of a model for interpretation. We emphasize that the quantitative interpretation will not depend on the assumptions used to reconstruct the image. Instead, $\mathcal{R}$ will be directly approximated, as described in Section~\ref{subs:DPVapprox}.

\section{DUST ENVELOPE}
\label{sec:model}

\subsection{Interpretation approach}

In this section, we interpret the observational data set under consideration. First, we analyze only SED to obtain an initial estimate of the symmetrical part of the envelope. Next, we will incorporate resolved scattered radiation and  construct a full 3D model of the source.

The observational data is represented by a vector $\boldsymbol{\mathrm{D}}$ of size $m$. The quantitative parameters of the object are denoted by the vector $\boldsymbol{\mathrm{\theta}}$. The purpose of the interpreting $\boldsymbol{\mathrm{D}}$ is to obtain quantitative estimates of $\boldsymbol{\mathrm{\theta}}$ and their uncertainties.

A crucial component of the interpretation is the physical model $\boldsymbol{\mathrm{M}}(\boldsymbol{\mathrm{\theta}})$, which enables calculation of the expected observations for a given $\boldsymbol{\mathrm{\theta}}$. We use numerical Monte--Carlo Radiation Transfer (MCRT) calculations, implemented in RADMC-3D software \citep{Dullemond2012}, as this model. RADMC-3D simulates the generation, propagation, scattering, and absorption of random discrete packets of photons (referred to here simply as ``photons''). The dependence of absorption coefficients and scattering matrices on wavelength and scattering angle is taken into account, along with changes in the polarization state of photons. Thermal equilibrium is maintained using the method of \citet{Bjorkman2001}. When a photon escapes the model space, it is collected. The parameters of the photons are then used to construct the SED and the image of the object, in all four Stokes parameters.

To organize the simulation and facilitate comparison with observational data we developed a web--based tool \href{https://circum.sai.msu.ru/}{webMCRT}. The analysis presented here was conducted using this system, and throughout the text, the readers will find direct links to specific models within it. A detailed description of webMCRT will be provided in a separate paper.

To estimate parameters, we use the maximum likelihood method. The residual is denoted as $\boldsymbol{\mathrm{R}}=\boldsymbol{\mathrm{D}}-\boldsymbol{\mathrm{M}}(\boldsymbol{\theta})$. The likelihood function is defined as follows:
\begin{equation}
p(\boldsymbol{\mathrm{D}}|\boldsymbol{\mathrm{\theta}}) = \frac{1}{\sqrt{(2\pi)^{m} \det(\mathrm{C})}} \exp\biggl(-\frac{1}{2} \boldsymbol{\mathrm{R}}^T \mathrm{C}^{-1}\boldsymbol{\mathrm{R}}\biggr),
\label{eq:likelihood}
\end{equation}
where $\mathrm{C}$ is the covariance matrix of observational uncertainties. The posterior probability for $\boldsymbol{\mathrm{\theta}}$ given observables $\boldsymbol{\mathrm{D}}$ is expressed using Bayes' theorem:
\begin{equation}
p(\boldsymbol{\mathrm{\theta}}|\boldsymbol{\mathrm{D}}) = \frac{p(\boldsymbol{\mathrm{D}}|\boldsymbol{\mathrm{\theta}}) p(\boldsymbol{\mathrm{\theta}})}{p(\boldsymbol{\mathrm{D}})},
\label{eq:BayesTheorem}
\end{equation}
where $p(\boldsymbol{\mathrm{\theta}})$ is the prior probability of parameters $\boldsymbol{\mathrm{\theta}}$. Here, we assume uniform distribution for all parameters. We will use $p(\boldsymbol{\mathrm{\theta}}|\boldsymbol{\mathrm{D}})$ to estimate the optimal values of $\boldsymbol{\mathrm{\theta}}$ and their credible intervals.

The covariance matrix $\mathrm{C}$ may be known, but in some cases (as shown in the examples below), indepedent information on $\mathrm{C}$ might not be available. In such cases, the observational uncertainties can be modeled by a random Gaussian process determined by a set of parameters $\boldsymbol{\theta}_N$. We then concatenate the vector $\boldsymbol{\theta}_N$ with the vector of astrophysical parameters $\boldsymbol{\theta}$ in form the total parameter vector: $\boldsymbol{\theta}_T$. The corresponding likelihood function then becomes:
\begin{equation}\label{eq:SEDlikelihood_covar}
\begin{split}
p(\boldsymbol{\mathrm{D}}|\boldsymbol{\mathrm{\theta}},\boldsymbol{\mathrm{\theta}}_N) = & \frac{1}{\sqrt{(2\pi)^{M} \det(\mathrm{C}(\boldsymbol{\mathrm{\theta}}_N))}}\cdot \\ & \cdot\exp\biggl(-\frac{1}{2} \boldsymbol{\mathrm{R}}^T \mathrm{C}(\boldsymbol{\mathrm{\theta}}_N)^{-1}\boldsymbol{\mathrm{R}}\biggr).
\end{split}
\end{equation}

Bayesian inference, which we use to obtain the posterior probability $p$ for the parameter vector $\boldsymbol{\theta}_T$ given observations $\boldsymbol{\mathrm{D}}$, naturally incorporates additional uncertainty related to the newly introduced parameters $\boldsymbol{\theta}_N$. Indeed, $p$ can be marginalized over the nuisance parameters $\boldsymbol{\theta}_N$, yielding the posterior probability for the astrophysical parameters alone.

This approach is particularly relevant for observations with unknown or underestimated noise, and its applications in astrophysics are rapidly expanding. For example, \citet{Czekala2015} used it to estimate stellar atmospheres parameters from high--resolution spectra; \citet{Matesic2024} applied it to approximate transits in Kepler photometry; and \citet{Wolff2017} modeled noise in HST images of a protoplanetary disk.

\begin{figure}[t]
    \centering
    \includegraphics[scale=0.53]{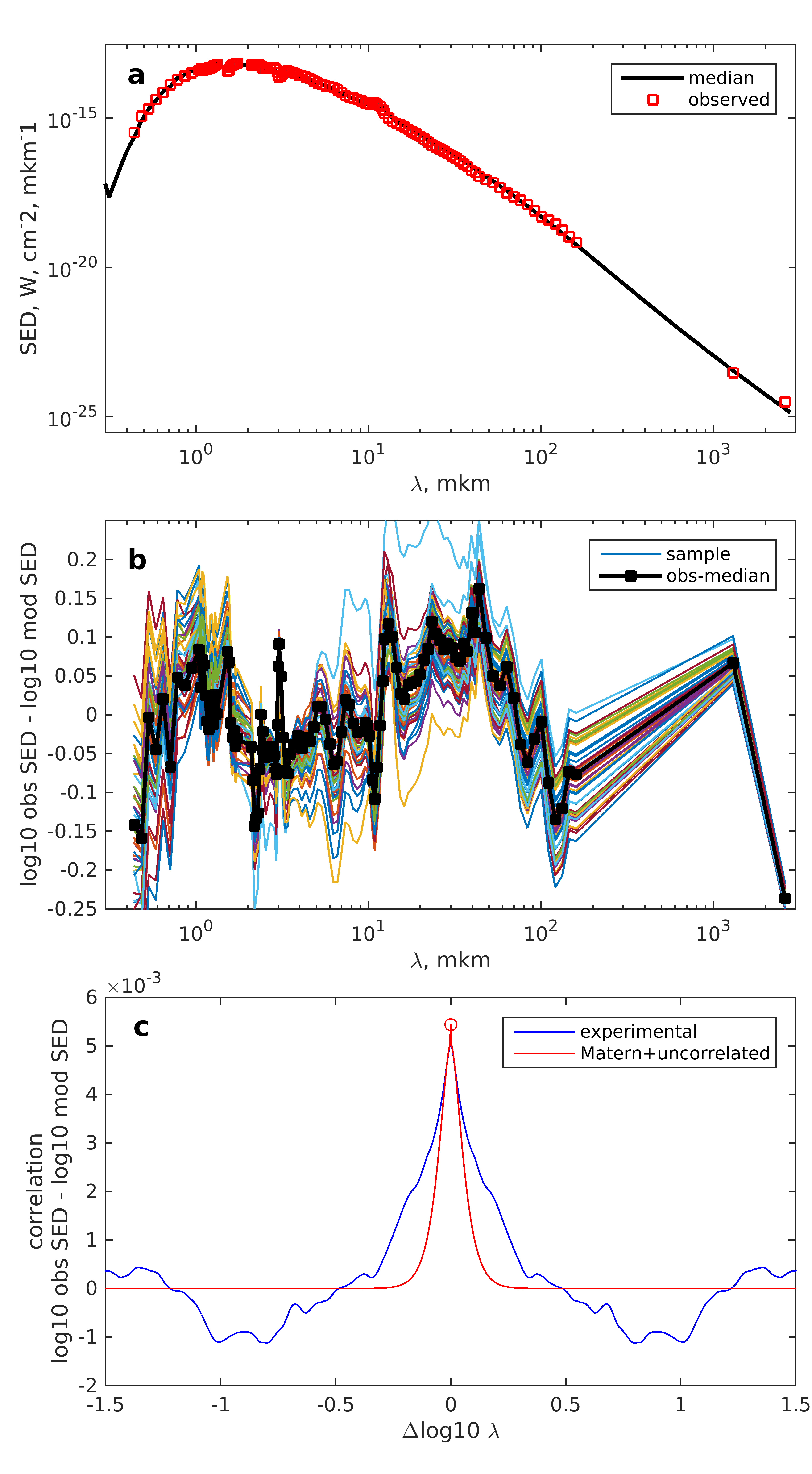}
    \caption{Approximation of SED. {\bf a)} Red squares represent the observed SED, while the black line shows the optimal model (corresponding to the median value of $\boldsymbol{\theta}$, as discussed in Section~\ref{subs:SEDapprox}). {\bf b)} Residuals $\boldsymbol{\mathrm{R}}$. Thick black line represents the optimal model, and the thin colored lines show a subsample of models drawn from the posterior distribution of $\boldsymbol{\theta}$. {\bf c)} Residual correlation. The blue line indicates the observed correlation, and the red line represents the model (\ref{eq:SEDcovar}) with parameters $\sigma_0=0.020$, $\sigma_c=0.71$, $l=0.060$,  as determined in Section~\ref{subs:SEDapprox}.}
    \label{fig:SEDapprox}
\end{figure}

\subsection{Interpretation of SED}
\label{subs:SEDmodel}

\subsubsection{Construction of likelihood function for SED}
\label{subs:SEDnoise}

While the internal relative uncertainties of the flux measurements used to construct the SED are quite small --- often $\lesssim1$\% --- several factors can distort the compiled SED. 

V~Cyg is a Mira variable, meaning its flux varies periodically. Moreover, the light curve is not precisely reproduced from cycle to cycle. In the visual wavelengths, changes can be as large as $2-4$~times. Flux measurements were taken at different phases and cycles of pulsation. Although efforts were made to account for this by reducing the SED to the moment of maximum brightness, this procedure introduces additional, unknown uncertainties. Flux measurements were conducted using apertures of varying sizes and different instruments, which means an unknown fraction of the flux may not have been accounted for. Additionally, some measurements were made using wide--band filters. 

These issues introduce additional noise into the SED, with the amplitude and correlation properties of this noise remaining unknown. The enormous range of wavelengths and fluxes considered here adds another layer of complexity. The wavelength spans 2.5 orders of magnitude, while the flux varies by 6 orders of magnitude. Based on physical intuition, one can assume that the relative uncertainty of flux measurements is at least comparable across different wavelengths. Consequently, the logarithm of the ratio of modeled and observed SED emerges as a convenient metric for assessing their deviation. Without loss of generality, we use the decimal logarithm for this purpose. Hereinafter, the decimal logarithm of the SED will be used as the measurement vector, $\boldsymbol{\mathrm{D}}$. 

The residual computed in this way is presented in Fig.~\ref{fig:SEDapprox}, b for the optimal model. The optimal model is defined as the one corresponding to the median values of the posterior distribution of $\boldsymbol{\theta}$ (see Section~\ref{subs:SEDapprox} for details). It is noteworthy that the optimal parameter values are weakly dependent on the adopted noise model; the noise model primarily affects the credible intervals for the parameters. It can be seen that the typical residual values are consistent across the considered wavelength range.

Another important property of the residual is that its values at adjacent wavelengths are strongly correlated. Moreover the correlation length remains nearly constant when the SED is considered as a function of the logarithm of the wavelength. For clarity, we use the decimal logarithm. The correlation function of the residual, $\boldsymbol{\mathrm{R}}$, is shown in Figure~\ref{fig:SEDapprox}.

We model the correlation of SED noise using the following approach, based on modified Matern kernel at $\nu=3/2$ \citep{Rasmussen2006}:
\begin{equation}
\mathrm{C}_{ij} = \sigma_0^2 \mathrm{I} + \sigma_c^2 \mathcal{K}_{ij},
\label{eq:SEDcovar}
\end{equation}
where $i,j$ are the indices of the measurement in the vector $\boldsymbol{\mathrm{D}}$, and $\mathrm{I}$ is the identity matrix. The elements of matrix $\mathcal{K}$ are calculated according to the rule:
\begin{equation}
\mathcal{K}_{ij} = \biggl( 1 + \frac{\sqrt{3}r_{ij}}{l}\biggr) \exp \biggl(-\frac{\sqrt{3}r_{ij}}{l} \biggr).
\label{eq:MaternKernel}
\end{equation}
Here, $r_{ij}$ is the distance between points, which we define as the decimal logarithm of wavelengths ratio. The parameters $\sigma_0^2, \sigma_c^2, l$ determine noise model, where $\sigma_0^2$ is the amplitude of uncorrelated noise, $\sigma_c^2$ is the amplitude of correlated noise, and $l$ is the correlation length. In Figure~\ref{fig:SEDapprox}, c we show how equation (\ref{eq:MaternKernel}) approximates our correlation function with $\sigma_0=0.020$, $\sigma_c=0.071$, $l=0.020$. These values were determined in the subsequent section.

\begin{deluxetable*}{ccccccc}
\caption{
Results of approximation of observations of SED and DPV using radiation transfer modeling.
\label{tab:sphspace}}
\tablewidth{0pt}
\tablehead{   & \multicolumn{2}{c}{spherical} & \multicolumn{2}{c}{spherical}  & \multicolumn{2}{c}{spherical + disk} \\
\colhead{parameter} & \colhead{grid}  & \colhead{max $p$.}      & \colhead{prior} & \colhead{optimal} & \colhead{prior} & \colhead{optimal}}
\colnumbers
\startdata
$f_\mathrm{C}$ & 0.65, 0.7, 0.75, 0.8, 0.85                    & 0.85     & $[0.285,0.923]$ &  $0.86\substack{+0.04 \\ -0.07}$    & $[0.285,0.923]$ & $0.85\substack{+0.04 \\ -0.02}$ \\
$a_\mathrm{max}$,~$\mu$m & 0.4, 0.5, 0.6, 0.7, 0.8, 0.9, 1.0, 1.2 & 0.8      & $[0.2,2.5]$  &  $0.65\substack{+0.29 \\ -0.22}$    & $[0.2,2.5]$     & $0.95\substack{+0.07 \\ -0.03}$ \\
$b$              & $-2.05,-2,-1.95$                       & $-2$     & $[-3.0, -1.0]$       &  $-1.97\substack{+0.03 \\ -0.04}$   & $-2$            & \\
$r_\mathrm{in}$,~AU  & 10, 12, 14, 16, 18, 20, 22             & 10       & $[10.0, 30.0]$   &  $11.4\substack{+2.2 \\ -1.0}$      & $10$            & \\
$\tau_\mathrm{sph}$  & 4.0                                    & 4.0      & $[1.0, 15.0]$    &  $3.59\substack{+0.28 \\ -0.28}$    & $[1.0, 15.0]$   & $2.97\substack{+0.06 \\ -0.27}$ \\
$\sigma_0$           &                                        &          & $[0.001,1.5]$    &  $0.02\substack{+0.004 \\ -0.003}$  & 0.020 & \\
$\sigma_c$           &                                        &          & $[0.001,1.5]$    &  $0.071\substack{+0.017 \\ -0.010}$ & 0.071 & \\
$l$                  &                                        &          & $[0.1,1.0]$      &  $0.060\substack{+0.023 \\ -0.013}$ & 0.060 & \\
$\tau_\mathrm{disk}$  &                                       &          &                  &                                                                & $[5, 60]$       & $33.0\substack{+9.4 \\ -4.0}$ \\
$h_0$,~AU                 &                                   &          &                  &                                                                & $[0.3, 4.0]$    & $1.81\substack{+0.66 \\ -0.26}$ \\
$\beta$                 &                                     &          &                  &                                                                & $[-3.0, 1.0]$ & $-1.33\substack{+0.23 \\ -0.72}$ \\
$\epsilon, ^{\circ}$                 &                        &          &                  &                                                                & $[50, 88]$      & $68.25\substack{+1.56 \\ -1.51}$ \\
\enddata
\tablecomments{
First column contains the model parameter: $f_\mathrm{C}$ --- the carbon fraction in the dust material, $a_\mathrm{max}$ --- the maximum radius of the dust particles, $\mu$m, $b$ --- the density exponent in the spherical envelope, $r_\mathrm{in}$ --- the inner radius of the envelope, AU, $\tau_\mathrm{sph}$ --- the optical depth in the spherical envelope at $\lambda=0.5$~$\mu$m, $\sigma_0$ --- the amplitude of uncorrelated noise in the SED, $\sigma_c$ --- amplitude of correlated noise in the SED, $l$ --- the correlation length of SED noise, $\tau_\mathrm{disk}$ ---  the optical depth in the disk at $\lambda=0.5$~$\mu$m, $h_0$ --- the disk scale height at 10~AU, $\beta$ --- exponent of change of the disk scale height variation with the distance from the star, $\epsilon$ --- the disk inclination, $^{\circ}$. The second and third column contain the grid and optimal values for approximation of SED by the model of spherical envelope using grid search. Fourth and fifth columns contain priors and optimal values of parameters for approximation of SED by the model of spherical envelope using MCMC sampling, Section~\ref{subs:SEDapprox}. Sixth and seventh columns contain priors and optimal values of parameters for approximation of SED and DPV by the model of spherical envelope and disk using MCMC sampling, Section~\ref{subs:DPVapprox}.}
\end{deluxetable*}

\subsubsection{Model of spherical envelope}
\label{subs:SEDapprox}

As a first order approximation of the SED we considered model of a spherical envelope. The SED noise covariance matrix was parameterized according to Equation~(\ref{eq:SEDcovar}). We assumed spherical particles composed of amorphous carbon and silicon carbide, with the mass fraction of carbon denoted as $f_\mathrm{C}$. The optical properties were adopted from \citep{Suh2000} and \citep{Pegourie1988} for carbon and SiC, respectively. The dust particle size distribution follows a power law with a fixed exponent $-3.5$, with the minimum radius fixed at 0.005~$\mu$m. The maximum radius, $a_\mathrm{max}$, is a parameter of the model. Parameters of absorption and scattering matrices were calculated using Mie theory with the optool\footnote{https://github.com/cdominik/optool} software \citep{Dominik2021}. 

The density of the envelope depends on the distance to the star, $r$, as:
\begin{equation}
\rho_\mathrm{sph} = \rho_{0,\mathrm{sph}} r^b.
\end{equation}
The outer radius was fixed at $r_\mathrm{out}=30000$~AU, as its exact value is poorly constrained by our observational data. The inner radius, $r_\mathrm{in}$, cannot be smaller than $10$~AU, as dust temperatures at closer distances  exceed sublimation temperature of 1400~K. The normalization factor $\rho_{0,\mathrm{sph}}$, and consequently the total mass of dust in the envelope, was defined by optical depth $\tau_\mathrm{sph}$ at $\lambda=0.5$~$\mu$m. 

Therefore we have the following vector of astrophysical parameters: $f_\mathrm{C}, a_\mathrm{max}, b, r_\mathrm{in}, \tau_\mathrm{sph}$. For the preliminary search for the global maximum of $p$, we assumed a simplified model for $\boldsymbol{\mathrm{C}}$: $\sigma_0=0.1$, $\sigma_c=0.0$. In other words, we assumed uncorrelated relative error of 10\%. This simplification will later be abandoned when optimizing the model using the Markov Chain Monte Carlo (MCMC) method.

RADMC-3D enables the computation of radiation transfer, accounting for both the absorption of stellar radiation and thermal dust radiation. Additionally, it allows for inclusion of scattering at various levels of approximation. Detailed calculations using scattering matrices result in changes to the resulting SED by $10-30\%$ in visible wavelengths compared to the case where scattering is neglected entirely. Clearly, accounting for scattering is essential for obtaining unbiased estimates of envelope parameters. However, such detailed scattering calculations increase computation time by $\approx50$ times, rendering the sampling of large parameter space computationally intractable.

To accelerate the calculation of the SED with scattering, we computed a correction factor $\xi(\lambda)=\mathrm{SED}_\mathrm{scat}(\lambda)/\mathrm{SED}_\mathrm{noscat}(\lambda)$ at nodes of a coarse grid of parameters, see Appendix~\ref{app:SEDcorr}. The dependencies $\xi(\lambda)$ are presented in Figure~\ref{fig:SED_correction_sph}. As one can see, the correction is largest in the visible wavelengths and exhibits minimal variation across different points of parameters space. When modeling the SED for a given parameter set ($f_\mathrm{C}, a_\mathrm{max}, b, r_\mathrm{in}, \tau_\mathrm{sph}$), the SED was calculated without scattering, and the resulting values were multiplied by the correction factor $\xi(\lambda)$, interpolated for the parameters of the given model. The precision of this correction is discussed at the end of this section.

Table~\ref{tab:sphspace} provides the parameter grid used for the search of the global maximum of $p$, as well as the parameters corresponding to this maximum. A comparison of the observed and modeled SED is presented in Figure~\ref{fig:SEDapprox} a.

After determining the approximate position of the global maximum of $p$, we employed MCMC approach to refine the position of the maximum and to estimate its credible intervals. Uniform distributions, as specified in Table~\ref{tab:sphspace}, were adopted as priors for all parameters. The MCMC procedure utilized 32 walkers, each producing 2500 samples of $p$. Convergence was achieved after the first $25\%$ of the samples, and remaining samples were used to construct the posterior distribution of $\boldsymbol{\theta}$\footnote{The chain evolution is presented on the \href{https://circum.sai.msu.ru/webMCRT/mcmcviewer?model_id=6735ee8aea54f5bfd61e681f&study_id=6736fb0bb4da446522c926d5}{webpage}}. Figure~\ref{fig:SEDcorner} in Appendix~\ref{app:corner} illustrates the posterior distribution. The optimal parameters and their credible intervals, defined as the 16-th and 84-th percentiles, are provided in Table~\ref{tab:sphspace}. The optimal SED is shown in Figure~\ref{fig:SEDapprox} a, and its deviation from the observed SED is represented by the green line in the panel b of the same figure. The reduced $\chi_r^{2}$ for the optimal model is 0.954.

Some weak correlations among the astrophysical parameters were observed. Models with larger particles radii $a_\mathrm{max}$ tend to have smaller optical depths $\tau_\mathrm{sph}$. Similarly, a larger inner radius of the envelope $r_\mathrm{in}$ corresponds to the smaller density exponent $b$. No significant deviation of $b$ from $-2$ was found, indicating that the material distribution in the envelope is consistent with the hypothesis of constant-velocity outflow and a steady mass loss rate. Furthermore, there is no evidence to suggest that the inner edge of the envelope extends farther from the star than the sublimation radius.

\begin{figure*}
  \center
  \includegraphics[width=0.85\linewidth]{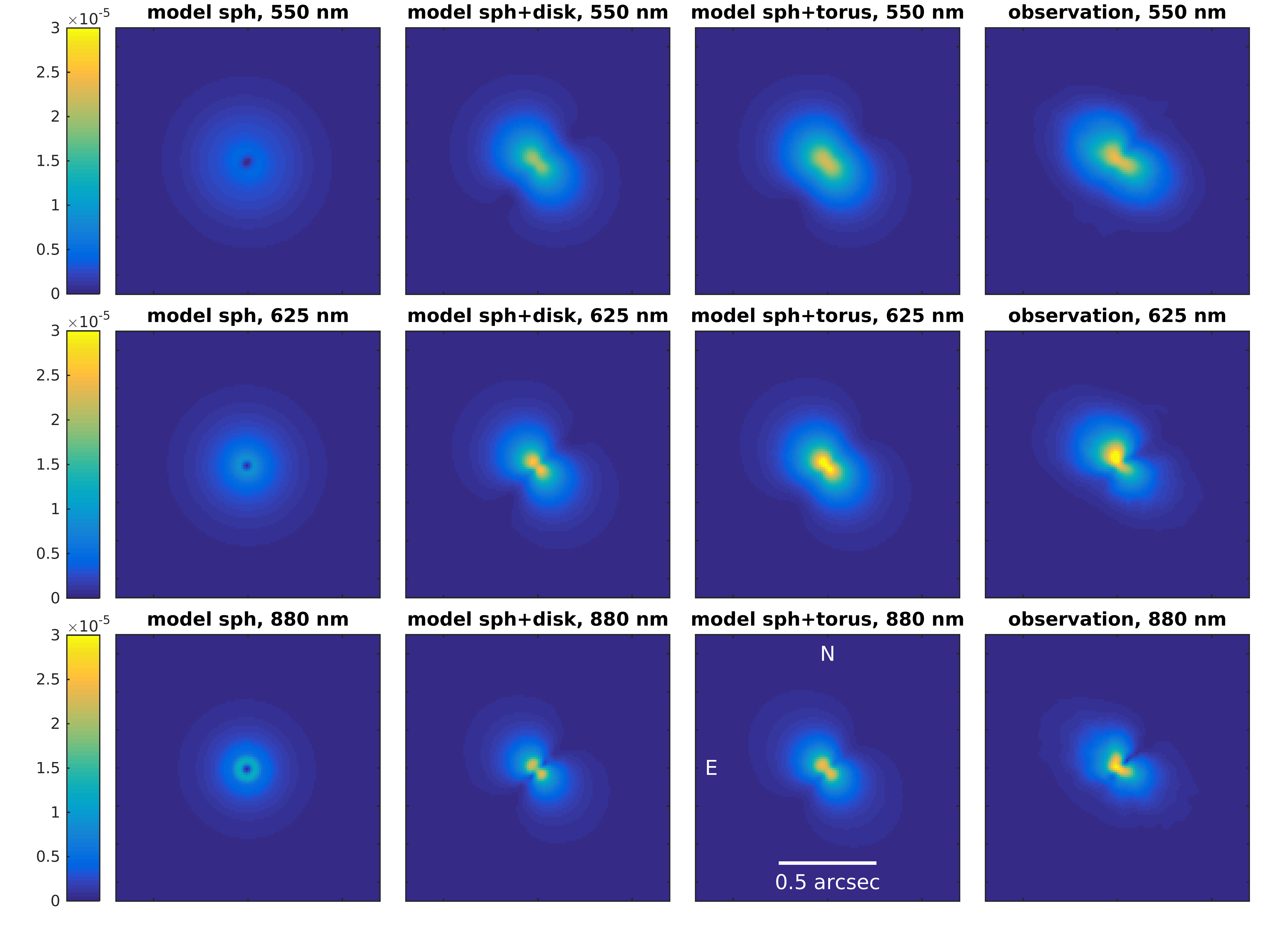}
  \caption{Comparison of observed (right column) and modeled (left and centered column) images of the V Cyg dust envelope in polarized intensity. The first column corresponds to the model of a spherical envelope (Section~\ref{subs:SEDmodel}). The second column corresponds to the model of a spherical envelope and disk (Section~\ref{subs:DPVmodel}). The third column corresponds to the model of a spherical envelope and torus (Section~\ref{subs:DPVmodel}). The brightness units in the color scale represent polarized flux in a $20\times20$ mas pixel relative to the total flux from the star.}
  \label{fig:images_comp}
\end{figure*}

To validate the method of SED correction for scattering, we selected a subsample of 50 realizations from the ensemble  generated by the MCMC. For this subsample, we calculated SEDs using two methods: the correction by interpolated $\xi(\lambda)$ as described above, and precise consideration of scattering. The ratio of SEDs computed using these two methods is presented in Figure~\ref{fig:SED_correction_sph}. As can be seen, the differences are minor, reaching only a few percent in the visible range. We conclude that the SED correction method is effective, providing accurate results, while significantly accelerating the calculations. The deviations of the SEDs in the subsample from the observed SED are presented in Figure~\ref{fig:SEDapprox}, b.

\subsection{Joint approximation of SED and DPV}
\label{subs:DPVmodel}

\subsubsection{DPV in the model of spherical envelope}

We computed images of the spherical envelope in Stokes parameters $I, Q, U$ using RADMC--3D, based on the model constructed in the previous subsection. The dependence of scattering matrices on wavelength and scattering angle was fully accounted for; and multiple scattering was taken into account (the most detailed approach available in RADMC--3D). The resulting images were convolved with the PSF corresponding to spectral filter, weighted inversely proportional to the square of the DPV uncertainty, for details see Section~\ref{subs:img}. The final polarized intensity distributions are shown in Fig.~\ref{fig:images_comp} in the left column. The dark spot at the center is not the inner edge of the envelope, its size is determined by the highest spatial frequencies attainable in Fourier space. The reduction of atmospheric turbulence effects and the higher available flux of the object at longer wavelengths result in a larger accessible region in Fourier space and, therefore, a smaller dark spot size.

\begin{figure*}
  \center
  \includegraphics[width=1.0\linewidth]{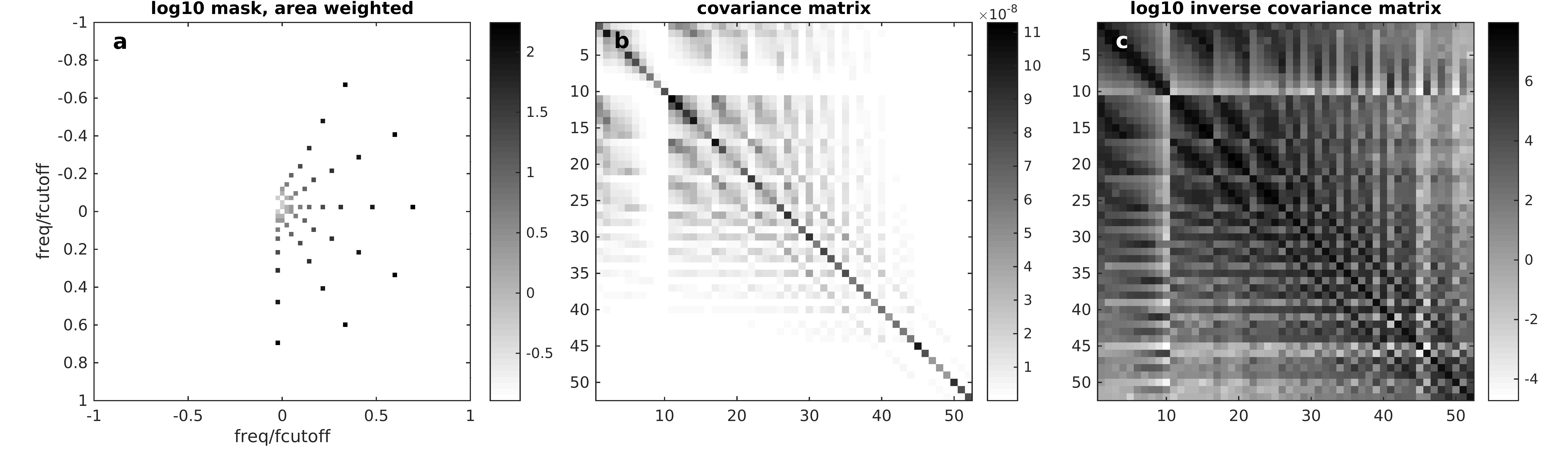}
  \caption{{\bf a)} Thinning mask in Fourier space, brightness is proportional to logarithm of weight. {\bf b)} Covariance matrix for DPV noise. {\bf c)} Logarithm of inverse of covariance matrix for DPV noise.}
  \label{fig:DSPnoiseCovMatrix}
\end{figure*}

The angular size of the modeled images roughly matches the observations (right column in Figure~\ref{fig:images_comp}). However, the overall brightness predicted by the spherical envelope model is significantly lower than observed. Furthermore, the relative brightness increases toward longer wavelengths, while in the observations the opposite trend is observed. Notably, the observations reveal a significant departure from central symmetry in the density distribution. The spherical envelope model thus requires refinement to simultaneously fit the SED and accurately predict the observed DPV at three wavelengths. In Section~\ref{subs:DPVapprox} we will construct such a refined model. For this, we will develop a likelihood function for the DPV data.

\subsubsection{Likelihood function for DPV}
\label{subs:DPVnoise}

To construct a likelihood function for DPV observations, we first investigate the noise characteristics of these observations. An example of a noise realization (N=18 harmonics, $\cos$--demodulation) for observations in the 880~nm band is shown in Figure~\ref{fig:DSPnoise}, a. In total, we have 60 such realizations, corresponding to different harmonic numbers. The most noticeable property of the noise is that its amplitude increases with the absolute value of the spatial frequency $|f|$. This behavior is expected, as the variance of the noise is inversely proportional to the average power spectrum of the image, which decreases with $|f|$ \citep{Safonov2023}. The dependence of the noise standard deviation on $|f|$ is shown in Fig.~\ref{fig:DSPnoise}, c.

Secondly, the noise is clearly correlated in Fourier space. The two--point correlation function is plotted in Figure~\ref{fig:DSPnoise}, b. At small separations between points in frequency space $f_r$, noise is almost fully correlated, with correlation decreasing as $f_r$ increases. The correlation function was computed in four concentric zones, as depicted in panel a of Figure~\ref{fig:DSPnoise}. The dependence of correlation on $f_r$ is similar across different frequency zones and can be approximated by a single exponential function $\exp(-f_r/f_l)$, where $f_l=0.065f_c$. This frequency scale corresponds to an angular scale of $1.1^{\prime\prime}$ in the image plane, which is comparable to the seeing $\beta$.

To compute the likelihood, the measurements of $\mathcal{R}_Q$ and $\mathcal{R}_U$ need to be unfolded into a vector. The covariance matrix for this vector will be modeled as follows:
\begin{equation}
\mathrm{C}_{ij}^{\mathrm{DPV}} = \sigma_i \sigma_j \exp{\bigl(-r_{ij}/f_l\bigr)},
\end{equation}
where $\sigma_i, \sigma_j$ are the noise standard deviation at points $i$, $j$, and  $r_{ij}$ is the distance between these points in Fourier space. The parameter $f_l$, previously defined, represents the correlation scale in Fourier space. Since all these quantities are known, $\mathrm{C}_{ij}^{\mathrm{DPV}}$ can, in principle, be evaluated. However, in practice, this poses significant challenges due to the large size of the covariance matrix. The number of measurements for $\mathcal{R}$ is $\approx10^4$ per one photometric band, resulting in covariance matrix of size $10^4\times10^4$. Computing the inverse of such a matrix, a necessary step for the likelihood (\ref{eq:SEDlikelihood_covar}) function, exceeds the available computational resources.

Despite this, it is evident that the measurements, similar to presented in Figure~\ref{fig:DSPobs}, have far fewer degrees of freedom than $10^4$. As we demonstrated in Figure~\ref{fig:DSPnoise}, the noise of $\mathcal{R}$ is correlated, meaning the measurements can be thinned in Fourier space using a step size corresponding to the correlation radius $f_l$. A similar approach was previously employed by \citet{Wolff2017} in image space. However, modifications are required for application in Fourier space. While the noise has a finite correlation radius, the signal may exhibit variations at any scale, particularly at low spatial frequencies. This differs from image space, where the support of the signal's Fourier spectrum is limited by the cut--off frequency $D/\lambda$.

The critical information about the object morphology at large angular scales, including those larger than seeing, is contained in the low--frequency region of $\mathcal{R}$ measurements. These angular scales are constrained by the field of view, which is $2-3^{\prime\prime}$ in our case. To address this, we applied thinning to the $\mathcal{R}$ measurements, taking into account this factor. Measurements in the Fourier space were retained at distinct points, with the distances between points linearly increasing with the absolute value of the spatial frequency. This approach ensures that low-frequency regions, which carry crucial morphological information, are sampled more densely. The resulting sampling pattern is illustrated in the left panel of Figure~\ref{fig:DSPnoiseCovMatrix}. To optimize computational efficiency, we considered measurements corresponding to only one half-plane, as it contains all necessary information due to symmetry. Each retained measurement was assigned a weight proportional to the area it ``substitutes'' in the Fourier space.

Using this thinned dataset $\boldsymbol{\mathrm{D}}$ and its associated covariance matrix, we computed log-likelihoods $\log L_{550}$, $\log L_{625}$, $\log L_{880}$ for the bands at 550, 625, and 880~nm, respectively. To find the model that best approximates both the SED and DPV, we maximized the total log-likelihood:
\begin{equation}
    \log L = W_\mathrm{SED} \log L_\mathrm{SED} + \log L_{550} + \log L_{625} + \log L_{880}.
\end{equation}
This formulation assumes statistical independence of noise realizations in the SED and DPV measurements across all three bands. To balance the sensitivity between SED and DPV data, we assigned a weight of $W_\mathrm{SED}=3$. For the computation of $\log L_\mathrm{SED}$, we used a covariance matrix parameterized with $\sigma_0=0.020$, $\sigma_c=0.071$, $l=0.060$, see subsection~\ref{subs:SEDnoise}. 

\subsubsection{Model of spherical envelope and disk}
\label{subs:DPVapprox}

The ``shadows'' in the image of the envelope (Figure~\ref{fig:images_comp}, right column) at PA$\approx135^{\circ}$ and $\approx315^{\circ}$ indicate deviations in the dust distribution from central symmetry. As a first-order approximation of this asymmetry, we considered axially symmetric distributions that also exhibit a plane of symmetry perpendicular to the symmetry axis passing through the star. Specifically, we modeled the distribution as a combination of a spherical envelope and either a disk or a torus.

\begin{figure*}
  \center
  \includegraphics[width=0.8\linewidth]{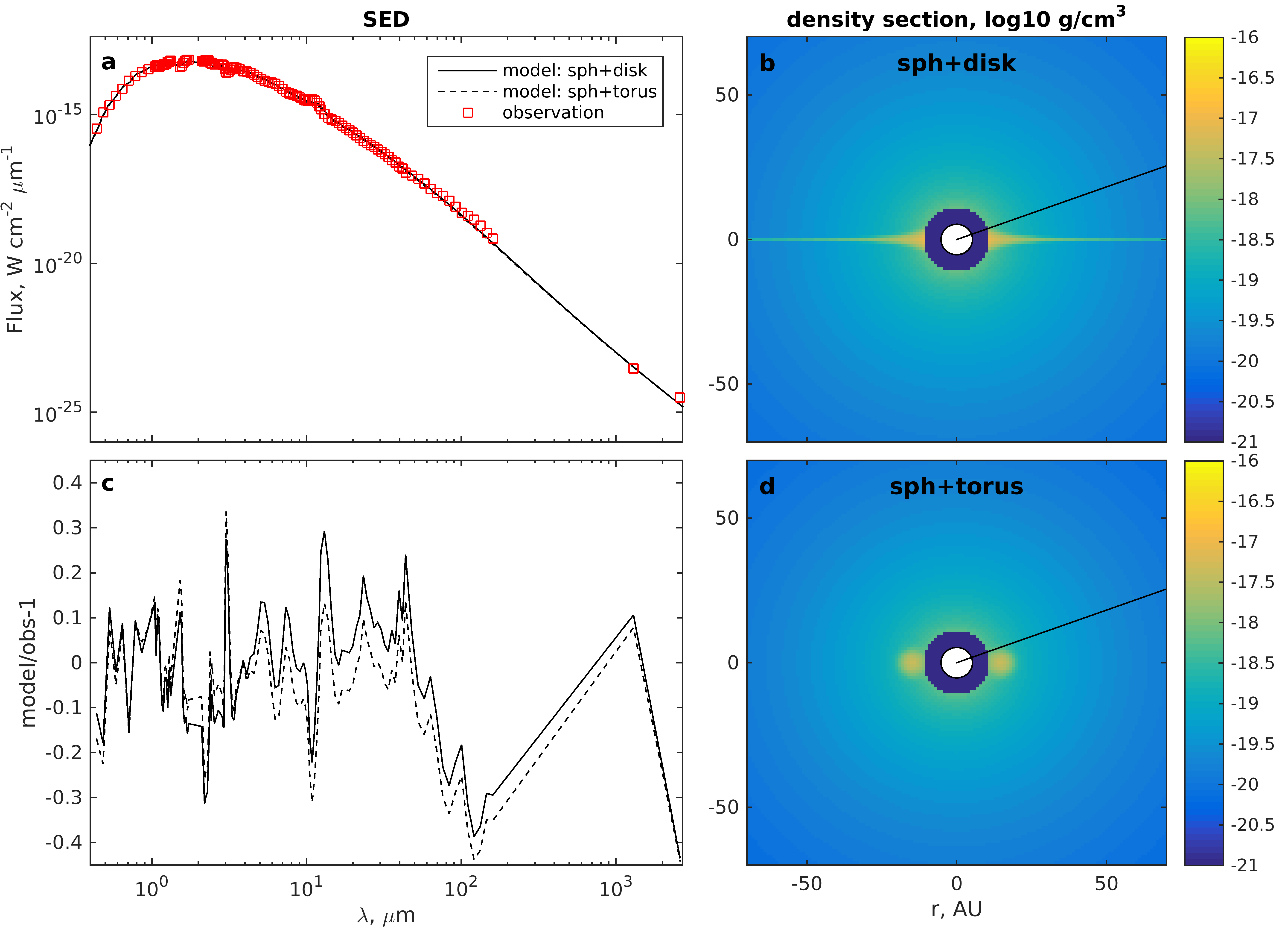}
  \caption{Models of spherical envelope with disk and spherical envelope with torus. a). SED. b): Decimal logarithm of density section for model with disk, white disk represents the star, black line --- line of sight. c). Relative deviation of the model SED from the observed SED. d). Same as b) but for model with torus. 
  \label{fig:SED_density}}
\end{figure*}

The combination of a spherical envelope and a disk has been employed in previous studies to explain the  morphology of protoplanetary nebulae such as IRAS~17441--2411, IRAS~08005--2356 \citep{Oppenheimer2005}, and M 1-92 \citep{Murakawa2010}. The presence of a disk is often attributed to the influence of a planetary or stellar companion. Similarly, the morphologies of mature planetary nebulae are frequently interpreted with the help of confining disks at small stellocentric distances~\citep{Decin2021}.

Following \citep{Oppenheimer2005, Murakawa2010}, we adopted a disk in hydrostatic equilibrium \citep{Shakura1973} for the model of dust density distribution (added to spherical envelope):
\begin{equation}
\rho_\mathrm{disk} = \rho_{0,\mathrm{disk}} r_{xy}^{-2} \exp \Bigl( -\frac{z^2}{2h^2(z)} \Bigr),
\end{equation}
\begin{equation}
h(z) = h_0 \Bigl( \frac{r_{xy}}{r_0} \Bigr)^{\beta},
\end{equation}
\begin{equation}
r_{xy}=\sqrt{x^2+y^2}.
\end{equation}
Here, $\rho_{0,\mathrm{disk}}$ is a normalization coefficient determined by the optical depth $\tau_{\mathrm{disk}}$ in the equatorial plane of the disk. The parameter $h_0$ represents the scale height of the disk at an arbitrarily set reference stellocentric distance $r_0=10$~AU, $\beta$ is characterizing the flaring of the disk. We assumed that the disk consists of the same dust as the spherical envelope and extends across the same distances from the star: from 11 to 30000~AU. The parameters of the spherical envelope were fixed to their optimal values determined in the previous section, except for the optical depth $\tau_\mathrm{sph}$ and the carbon fraction $f_\mathrm{C}$. The position angle of the disk's symmetry axis was fixed at $45^{\circ}$, based on the observed appearance of the envelope. In this model, the envelope's appearance also depends on the inclination $\epsilon$, which is $0^{\circ}$ if we are looking from the axis of the disk and $90^{\circ}$ if we are looking from the equator.

To constrain the parameters $\tau_{\mathrm{sph}}, f_\mathrm{C}, \tau_{\mathrm{disk}}, h_0, \beta, \epsilon$ we employed MCMC. The parameter space was sampled by 32 walkers, each performing 1830 iterations to measure the probability density $p$. The chains reached convergence after approximately 700 iterations\footnote{The chain evolution is presented on the \href{https://circum.sai.msu.ru/webMCRT/mcmcviewer?model_id=67334bcf6ce7853c1cda5fb6&study_id=674727a4f4af2a57db02f6d4}{webpage}}. For the construction of the posterior distributions, 60\% of the probability measurements were used. The results are presented in Figure~\ref{fig:SEDDPVcorner}, and the priors, optimal parameter values and $1\sigma$ uncertainties are summarized in columns 6 and 7 of Table~\ref{tab:sphspace}.

The corresponding SEDs and the density cross-section of the model are presented in Figure~\ref{fig:SED_density}. For the SED $\chi_r^2=1.01$, indicating a good fit. The total masses of the spherical envelope and the disk were determined to be $5.3\times10^{-5} M_\odot$ and $2.3\times10^{-8} M_\odot$ ($\equiv7.6\times10^{-3} M_\oplus$), respectively.

The images in polarized intensity at wavelengths 550, 625, and 880~nm are shown in the second column of Figure~\ref{fig:images_comp} . The model that includes a disk explains several observed features: the shadows at PA$\approx135^{\circ}$ and $\approx315^{\circ}$, the difference in brightness between the northeast and southwest lobes and overall dependence of envelope brightness on the distance from the star. The reduced chi-squared values $\chi_r^2$ for DPV measurements are 0.72, 2.21, 4.41 in the 550, 625, 880~nm bands, respectively. These results indicate that the model combining a spherical envelope and a disk provides consistent explanation for the entire set of observations. The somewhat elevated $\chi_r^2$ at 880~nm is likely due deviations in morphology from symmetry around the axis PA=$45^{\circ}$, which are not accounted for in the model. Additionally, observations in this band benefit from a higher signal-to-noise ratio, making deviations more apparent.

\begin{figure}[t!]
  \center
  \includegraphics[width=1.0\linewidth]{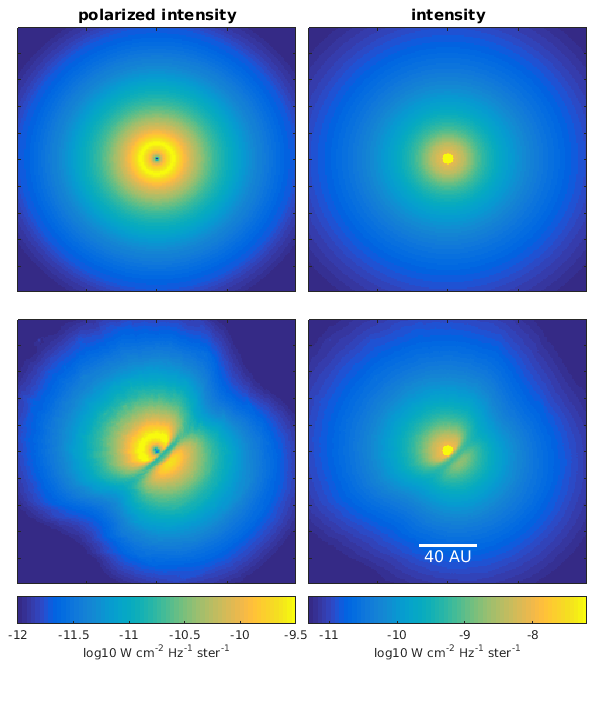}
  \caption{The modeled images of the system at $880$~nm, not convolved by the PSF. Left column: the polarized intensity, right column: the total intensity. Upper row: the model of the spherical envelope, Section~\ref{subs:SEDapprox}, lower row: the model of the spherical envelope and disk, Section~\ref{subs:DPVapprox}.}
  \label{fig:raw_simul_image}
\end{figure}

The model is further illustrated in Figure~\ref{fig:raw_simul_image}, which shows the central parts the circumstellar environment at $\lambda=880$ without convolution with PSF and at a higher spatial resolution than in Figure~\ref{fig:images_comp}. The images compare the models with and without a disk in both total intensity and polarized intensity. In the polarized intensity images, the inner edge of the envelope, located at the sublimation radius, is distinctly visible. These images resolve the stellar disk, which is apparent in the total intensity images. In the case of the model that includes a disk, the shadow cast by the disk is clearly visible in both total and polarized intensity.

Note (Fig.~\ref{fig:SED_density}, b) that our modeling strongly favors negative $\beta$ values; in other words, the disk tapers towards its edges.  This configuration appears counterintuitive for a decretion disk in hydrostatic equilibrium. In Appendix~\ref{app:disks_beta} we further demonstrate that density distribution for $\beta=0$ (a disk of constant thickness) and $\beta=1$ (a flaring disk) predict results which are incompatible with observations in scattered light. The tapered structure of the disk results in almost all of its mass being concentrated at distances closer than 25~AU from the star. Consequently, even a very modest dust mass in the disk is sufficient to significantly alter the illumination pattern of the envelope. This peculiar result motivated us to consider a model with equatorial density enhancement of a different shape.

\subsubsection{Model of spherical envelope and torus}
\label{subs:torusapprox}

We considered an alternative geometry for the dust distribution with the same symmetry properties as the disk: a torus. The density of the torus is defined as follows:
\begin{equation}
\rho_\mathrm{torus}=\rho_{0,\mathrm{torus}} \exp \Biggl( -\frac{\sqrt{z^2+{(r_{xy}-r_\mathrm{ma})}^2}}{\sigma_\mathrm{torus}^2} \Biggr).
\end{equation}
Here the normalizing coefficient $\rho_{0,\mathrm{torus}}$ is defined by the optical depth $\tau_\mathrm{torus}$ in the equatorial plane of the torus. $r_\mathrm{ma}$ is the major radius of the torus, $\sigma_\mathrm{torus}$ defines its geometrical thickness.

The parameters of the torus were determined using MCMC fitting, similar to the approach used for the disk\footnote{The chain evolution is presented on the \href{https://circum.sai.msu.ru/webMCRT/mcmcviewer?model_id=6724c8a73df6d27c4ec71fdf&study_id=672a00808aeeed80c0689458}{webpage}}. The results are presented in Table~\ref{tab:sphtorusspace}.
The parameters of the spherical envelope are very similar to those determined for disk model. The total mass of the torus is $5.7\times10^{-3}$ M$_\oplus$.

\begin{deluxetable}{ccc}
\caption{Results of approximation of observations of SED and DPV by the model of spherical envelope and torus using radiation transfer modeling.
\label{tab:sphtorusspace}}
\tablewidth{0pt}
\tablehead{  \colhead{parameter} & \colhead{prior} & \colhead{optimal}}
\colnumbers
\startdata
$f_\mathrm{C}$           & $[0.285,0.923]$ & $0.88\substack{+0.04 \\ -0.05}$ \\
$a_\mathrm{max}$,~$\mu$m & $[0.2,2.5]$     & $0.99\substack{+0.28 \\ -0.06}$ \\
$b$                      & $-2$            & \\
$r_\mathrm{in}$,~AU      & $10$            & \\
$\tau_\mathrm{sph}$      & $[1.0, 15.0]$   & $2.83\substack{+0.10 \\ -0.16}$ \\
$\tau_\mathrm{torus}$    & $[2, 12]$       & $8.8\substack{+1.7 \\ -1.6}$ \\
$r_\mathrm{ma}$,~AU      & $[10.0, 25.0]$    & $14.7\substack{+1.22 \\ -1.94}$ \\
$\sigma_\mathrm{torus}$  & $[0.5, 10.0]$   & $2.17\substack{+0.18 \\ -0.22}$ \\
$\epsilon, ^{\circ}$     & $[50, 88]$      & $66.6\substack{+1.0 \\ -2.8}$ \\
\enddata
\tablecomments{
First column contains the model parameter: $f_\mathrm{C}$, $a_\mathrm{max}$, $\mu$m, $b$, $r_\mathrm{in}$, $\tau_\mathrm{sph}$ were described before in caption of Table~\ref{tab:sphspace}. $\tau_\mathrm{torus}$ ---  the optical depth in the equatorial plane of the torus at $\lambda=0.5$~$\mu$m, $r_\mathrm{ma}$ --- the major radius of the torus, $\sigma_\mathrm{torus}$ --- parameter defining thickness of the torus, $\epsilon$ --- the disk inclination, $^{\circ}$. Second and third columns contain priors and optimal values of parameters for approximation of SED and DPV by the model of spherical envelope and torus using MCMC sampling, Section~\ref{subs:torusapprox}.}
\end{deluxetable}

The model's density and the resulting SED are illustrated in Fig.~\ref{fig:SED_density}, with a reduced chi-squared value of $\chi_r^2=1.19$. The images in polarized intensity are given in Fig.~\ref{fig:images_comp}, third column. The reduced chi-squared values $\chi_r^2$ for the DPV measurements are 0.70, 1.98, 4.14 in the 550, 625, 880~nm bands, respectively. Despite the significant differences in dust geometry, models with a disk and with a torus produce very similar observables, including SED and DPV. They also predict similar dust density enhancement near the sublimation radius. In the equatorial plane at a distance of 15 AU from the star, the densities are $\rho=3.0\times10^{-18}$ g/cm$^3$ and $\rho=3.7\times10^{-18}$ g/cm$^3$ for the disk and the torus, respectively. This can be compared to the density of the spherical component, which is $\rho=4.4\times10^{-19}$ g/cm$^3$ and $\rho=4.2\times10^{-19}$ g/cm$^3$ at 15~AU from the star in the direction perpendicular to the equatorial plane for the models with disk and torus, respectively. The torus has lower optical depth in the equatorial plane with respect to the disk, which is related to weaker stratification of its material in the vertical direction. Based on the existing data, it is impossible to draw definitive conclusions about the properties of the equatorial density enhancement beyond 25 AU from the star.

\section{DISCUSSION}
\label{sec:discussion}

Direct estimation of the effective temperature $T_\mathrm{eff}$ of V~Cyg have been made using measurements of integrated flux and angular diameter by \citep{Bergeat2001,angsize}. \citet{angsize} determined $T_\mathrm{eff}=1949\pm167$ and $T_\mathrm{eff}=2051\pm174$ at pulsation phases $\phi=0.25$ and $\phi=0.81$, respectively. Using our NIR low-resolution spectra and updated data from the literature, we revised the integral flux, resulting in a new estimate of the effective temperature: $T_\mathrm{eff}=2600$~K.

This study highlights how constructing a model that consistently describes both the thermal and scattered radiation of circumstellar dust can resolve degeneracies that arise when only one data source is analyzed. The geometry of the envelope affects SED only in extreme cases, such as when all the material is concentrated in a thin disk, as observed around some young stars. More subtle deviations from central symmetry, such as those found in V~Cyg, leave negligible traces in the SED, which can sometimes be misinterpreted as a change in the mass-loss rate \citep{Wiegert2020}. Considering resolved scattered radiation is indispensable for accurately assessing the geometry of the envelope. While resolved thermal radiation could also help in this regard, achieving the necessary spatial resolution is more challenging. This is due to the longer wavelengths typical of thermal radiation from circumstellar dust, which demand larger apertures or extended interferometric baselines. 

In contrast, analyzing spatially resolved short-wavelength scattered radiation is more accessible, even with modest apertures. For instance, the 2.5-m telescope of CMO SAI MSU used in this study was sufficient to spatially resolve the envelope regions that dominate the thermal IR spectrum. An additional advantage of scattered radiation is its higher sensitivity to dust particle size compared to thermal radiation. In the case of V~Cyg, from Table~\ref{tab:sphspace} one can see that the upper limit of dust particles size distribution is determined with significantly greater precision --- several times more --- when DPV data is included alongside SED observations. 

From an observational perspective, the characterization of scattered radiation is significantly facilitated by its polarization, enabling the application of a wide array of polarimetric techniques. Polarimetric measurements are often more precise than corresponding direct flux measurements due to their differential nature, which suppresses many random and systematic errors \citep{Clarke2010}. Among these, polarimetric differential imaging (PDI) has emerged as a particularly effective method for reducing unpolarized stellar radiation in high-contrast observations \citep{Kuhn2001,Apai2004}. Over the past decade, PDI has revolutionized the study of protoplanetary disks, opening new avenues of research \citep{Garufi2024}. Similarly, measurements of differential polarimetric visibility, as demonstrated by \citet{Norris2012} and our own studies \citep{Safonov2019}, are $20-30\times$ more precise than measurements of total visibility. As noted in the introduction, resolving polarized radiation is essential for analyzing scattered radiation from circumstellar envelope, as integral polarization measurements are significantly compromised by averaging effects. In case of V~Cyg the combined use of resolved polarized data in the form of DPV and SED has enabled the discovery of new structures within  the innermost circumstellar environment and provided more precise mass-loss estimates.

Although axially symmetric structures resembling bipolar outflows constrained by disks are ubiquitous among planetary nebulae and proto--planetary nebula, they are relatively rare in less evolved AGB stars \citep{Decin2021}. One prominent example is the oxygen--rich AGB star L$_2$~Pup, where ALMA observations revealed a disk potentially shaped by a substellar companion \citep{Kervella2016,VandeSande2024}. Similarly, ALMA kinematic analysis of EP~Aqr circumstellar environment \citep{Nhung2024} revealed the an equatorial density enhancement attributed to a companion \citep{Homan2020}. 

The mass of the disk detected in this study is primarily confined to a region within 25~AU to the star. In this context, it is unlikely that the disk takes part in the outward motion of the material in the spherically symmetric envelope. An analysis of gas kinematics, similar to conducted by \citet{Bujarrabal2016}, is needed to determine whether a substellar companion could be responsible for the presence of the disk.

V~Cyg exhibits the highest water content in its envelope among known carbon stars. \citet{Neufeld2010}, through modeling emission lines observed with {\it Herschel}/HIFI, estimated a water mass-loss rate of $3-6\times10^{-4}$~M$_\oplus\times$yr$^{-1}$, which is $\approx10^4$ times higher than expected for stars of this type \citep{Cherchneff2006}. To account for this anomalously high water content in the envelopes of V~Cyg and other carbon stars, \citet{Neufeld2010,Neufeld2011} proposed the destruction of cometary bodies as a potential mechanism. In this scenario, water is entrained by the stellar wind at distances of 30--45~AU from the star. Taking into account that the water molecules are dissociated by interstellar UV radiation within a timescale of $30$~yr, the total  water content in the envelope of V~Cyg is estimated to be $1-2\times10^{-3}$$~$M$_\oplus$.

However, \citet{Neufeld2011} also observed that the profiles of water and hydrogen emission lines are very similar, a feature difficult to reconcile with the idea of water entrained from a disk. While several alternative hypotheses were proposed to explain presence of water in the stellar wind, none can quantitatively reproduce the observed water content \citep{Lombaert2016}. Notably, the mechanism involving cometary bodies destruction can produce water line profiles resembling those of hydrogen only if the observer's line of sight lies close to the equator of the disk. \citet{Neufeld2011} suggests that this mechanism might be plausible for some objects. Our study's detection of a dusty disk with a high inclination ($68^{\circ}$) aligns with this hypothesis. Additionally, the mass of dust in the V~Cyg's envelope ($7.3\times10^{-3}$~M$_\oplus$) is of the same order of magnitude as the estimated water content in its wind. The detection of a dusty disk along with high water content suggests the possibility of ongoing erosion of cometary bodies by stellar wind in V~Cyg.

\section{CONCLUSION}
\label{sec:conclusion}

We present measurements of infrared fluxes of carbon Mira V Cyg in $JKHLM$ bands and NIR spectra obtained at multiple epochs. These measurements, combined with literature data, were used to construct the SED of the source over the wavelength range $0.4-160~\mu$m for the moments of maximum and minimum brightness. 

To better constrain the dust properties in the envelope and study the spatial structure of V~Cyg's circumstellar environment, we conducted differential speckle polarimetry observations of the object at wavelengths 550~nm, 625~nm, and 880~nm. At these wavelengths, the dust in the envelope is expected to generate scattered polarized radiation. The envelope was clearly detected in all three bands at stellocentric distances of 50-300~mas. A significant departure of the envelope from central symmetry was found.

We modeled the source using 3D Monte Carlo radiation transfer simulation. The \text{models} of a spherical envelope with an equatorial density enhancement in the form of either thin disk or a torus successfully reproduce both the integrated thermal radiation and the resolved scattered light observations. The inner radius of the spherical envelope was determined to be $11.0\substack{+1.3 \\ -0.7}$~AU, corresponding to the sublimation distance. The outer radius, which is poorly constrained by our observations, was fixed at 30000~AU. The envelope's density is consistent with assumptions of constant expansion velocity and constant mass-loss rate. The total dust mass in the spherical envelope is estimated at $5.3\times10^{-5}$ M$_\odot$. Assuming a terminal velocity of 11.8 km$\cdot$s$^{-1}$ \citep{Neufeld2010} and a gas-to-dust mass ratio of 100:1, the mass loss rate is calculated to be $4.3\times10^{-7}$ M$_\odot\cdot$yr$^{-1}$.  

According to our model, dust particles radii follow power law with an exponent $-3.5$, ranging from $0.005~\mu$m to $0.95\substack{+0.07 \\ -0.03}~\mu$m. The dust material consists of amorphous carbon and silicon carbide (SiC), the content of amorphous carbon is $85\substack{+4 \\ -2}\%$.

In the model containing the disk, the latter shares the same inner and outer radii as the spherical envelope, with density decreasing with distance from the star as $r^{-2}$. The scale height of the disk decreases as $r^{-1.33\substack{+0.23 \\ -0.72}}$, leading to most of the disk mass being concentrated at stellocentric distances of $<25$~AU. The characteristic scale height is $1.81\substack{+0.66 \\ -0.26}$~AU at 10~AU. The optical depth along the equator of the disk is $33\substack{+9 \\ -4}$ at $\lambda=0.5$~$\mu$m, which corresponds to the total mass of $7.6\times10^{-3}M_\oplus$. The inclination of the disk is $68^{\circ}\substack{+2^{\circ} \\ -2^{\circ}}$.

As an alternative model of an equatorial density enhancement we have considered a torus with the Gaussian profile. The major radius of the torus was determined to be $14.70\substack{+1.22 \\ -1.94}$~AU, while its characteristic thickness is $2.17\substack{+0.18 \\ -0.22}$~AU. The optical depth along the equator is $\tau_\mathrm{torus}=8.8\substack{+1.7 \\ -1.6}$, yielding total dust mass of $5.7\times10^{-3}$ M$_\oplus$. We note that the morphology of the considered equatorial density enhancement, either in form of a disk or a torus, is incompatible with the assumption that it takes part in outward motion of the stellar wind.


We assumed that the disk/torus is composed of the same dust particles as the spherical envelope. Unfortunately, our observations do not provide the means to independently constrain the dust properties specific to the disk. We leave this for further studies, which should be based on high angular resolution imaging or interferometry in infrared and sub-mm wavelengths. Such analysis will help clarify the origin of the equatorial density enhancement and its intriguing connection to the high water content observed in the envelope of V Cyg.

\vspace{1.5cm}

We owe the engineering and scientific staff of Caucasian Mountain Observatory of SAI MSU. The comments by two anonymous referees allowed to substantially extend the analysis and improve the presentation. Authors acknowledge the support of M.V.Lomonosov Moscow State University Program of Development. S. Zheltoukhov acknowledges support of ``BASIS'' foundation of for development of theoretical physics and mathematics, project No. 21-2-10-35-1. The work of  A.Tatarnikov (spectral observations and data reduction) and V.Shenavrin (photometric observations and calibration) is supported by the RSF grant 23-22-00182. 

This publication makes use of data products from the Wide-field Infrared Survey Explorer, which is a joint project of the University of California, Los Angeles, and the Jet Propulsion Laboratory/California Institute of Technology, funded by the National Aeronautics and Space Administration.

The version of the ISO data presented in this paper correspond to the Highly Processed Data Product (HPDP) set called ``A uniform database of SWS 2.4-45.4 micron spectra'' by G.C. Sloan et al., available for public use in the ISO Data Archive

This publication makes use of data products from the Two Micron All Sky Survey, which is a joint project of the University of Massachusetts and the Infrared Processing and Analysis Center/California Institute of Technology, funded by the National Aeronautics and Space Administration and the National Science Foundation.

This research is based on observations with AKARI, a JAXA project with the participation of ESA.

This research made use of data products from the Midcourse Space Experiment. Processing of the data was funded by the Ballistic Missile Defense Organization with additional support from NASA Office of Space Science. This research has also made use of the NASA/ IPAC Infrared Science Archive, which is operated by the Jet Propulsion Laboratory, California Institute of Technology, under contract with the National Aeronautics and Space Administration.

This work has made use of data from the European Space Agency (ESA) mission {\it Gaia} (\url{https://www.cosmos.esa.int/gaia}), processed by the {\it Gaia} Data Processing and Analysis Consortium (DPAC, \url{https://www.cosmos.esa.int/web/gaia/dpac/consortium}). Funding for the DPAC has been provided by national institutions, in particular the institutions
participating in the {\it Gaia} Multilateral Agreement.

We acknowledge with thanks the variable star observations from the AAVSO International Database contributed by observers worldwide and used in this research.

%

\vspace{5mm}
\facilities{HST(STIS), Swift(XRT and UVOT), AAVSO, CTIO:1.3m,
CTIO:1.5m,CXO}


\software{astropy \citep{2013A&A...558A..33A,2018AJ....156..123A}}



\appendix



\section{Posteriors for envelope model}
\label{app:corner}

\begin{figure}[h!]
  \center
  \includegraphics[width=0.8\linewidth]{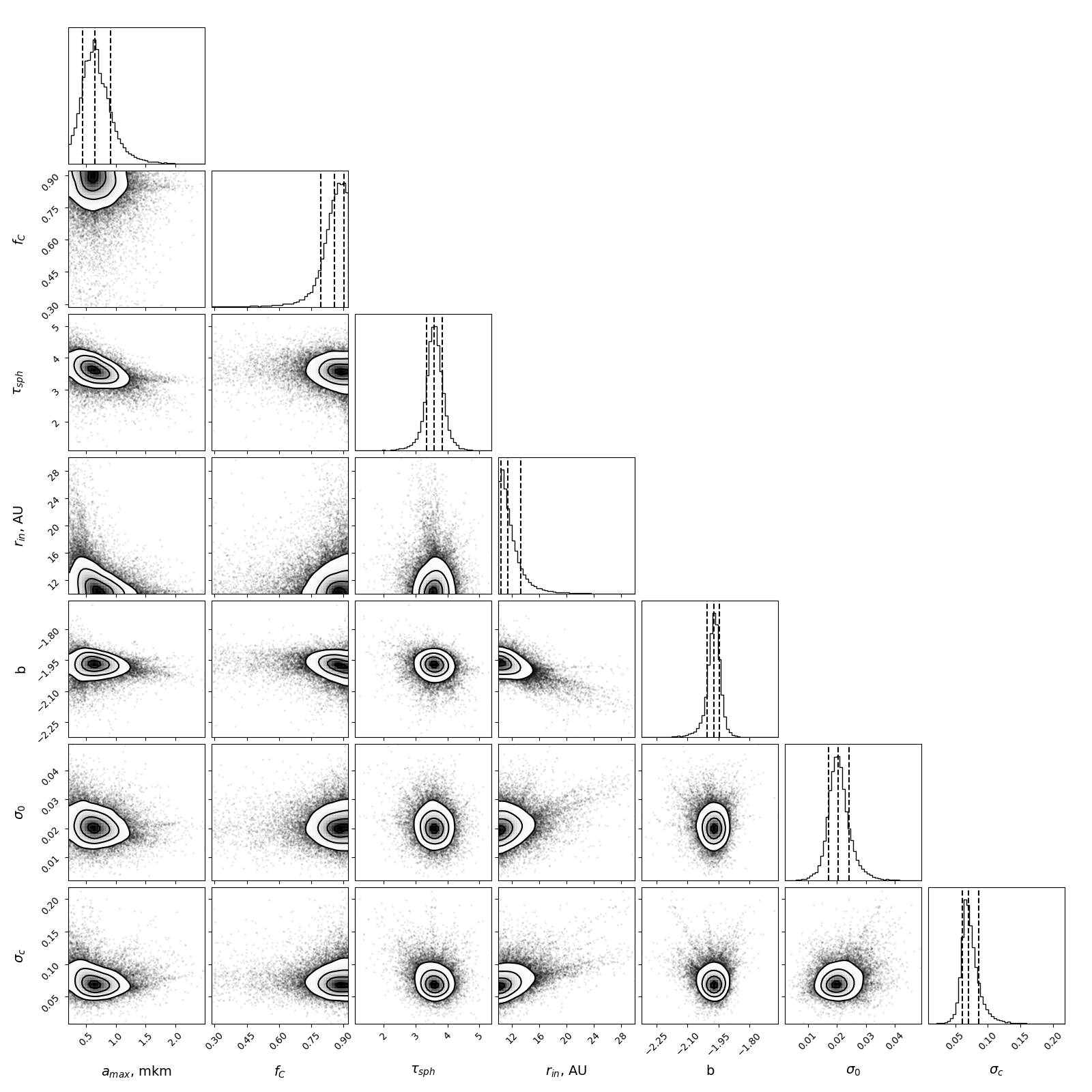}
  \caption{Corner plot of the posterior probability for parameters of the spherical envelope model, obtained using only the SED (Section~\ref{subs:SEDmodel}). Parameters: $a_\mathrm{max}$ --- maximum radius of dust particle, $f_C$ --- carbon fraction in dust material, $\tau_\mathrm{sph}$ --- optical depth in the envelope at $\lambda=0.5$~$\mu$m, $r_\mathrm{in}$ --- inner radius of envelope, $b$ --- exponent of envelope density. The rest three parameters characterize model of SED noise: $\sigma_0$ --- amplitude of uncorrelated noise, $\sigma_c$ --- amplitude of correlated noise, $l$ --- noise correlation length.}
  \label{fig:SEDcorner}
\end{figure}

\begin{figure}[h!]
  \center
  \includegraphics[width=0.8\linewidth]{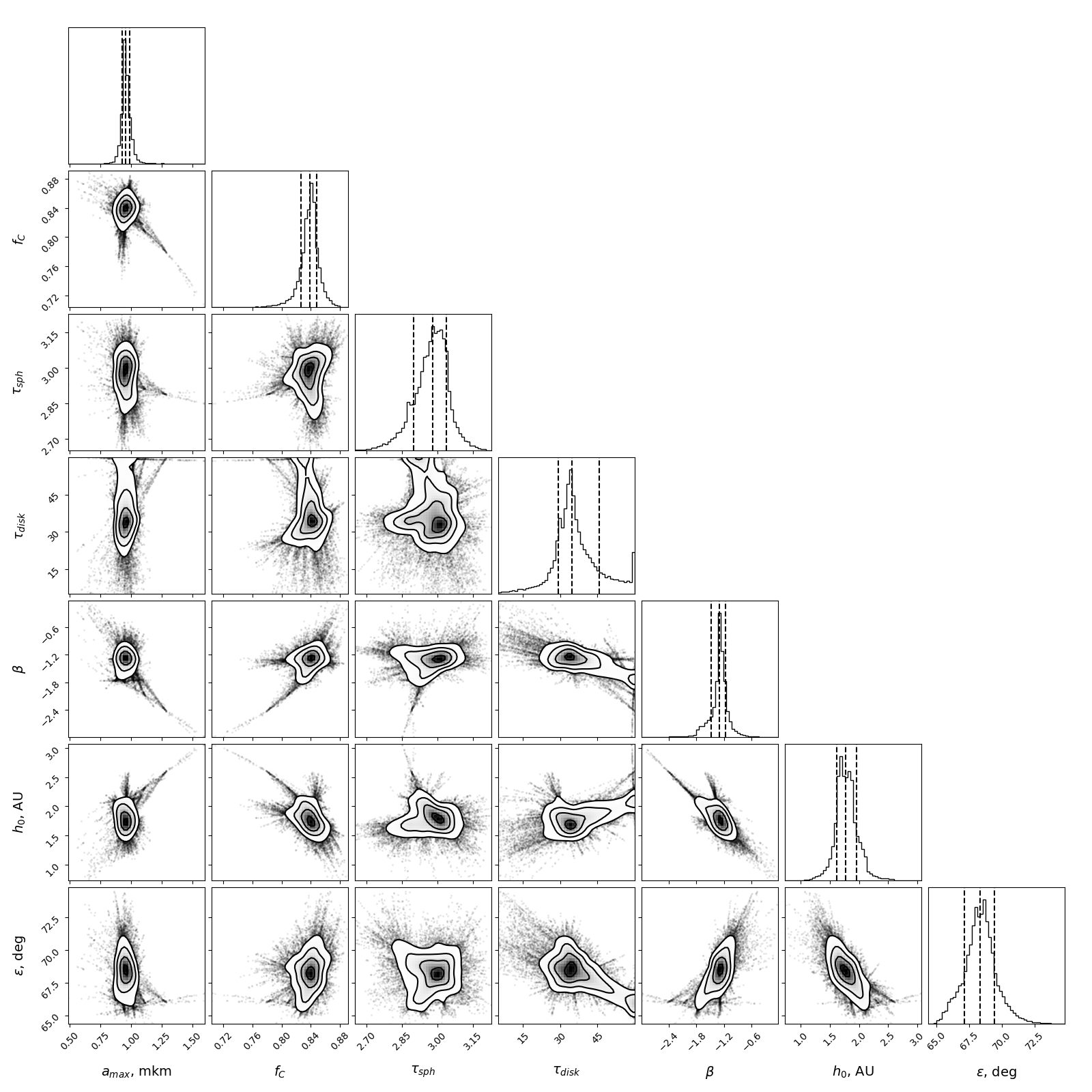}
  \caption{Corner plot of the posterior probability for parameters of the model of spherical envelope and disk, SED and DPV are taken into account (Section~\ref{subs:SEDmodel}). Parameters: $a_\mathrm{max}$ --- maximum radius of dust particles, $f_C$ ---  carbon fraction in dust material, $\tau_\mathrm{sph}$ --- optical depth in spherical envelope at $\lambda=0.5$~$\mu$m, $\tau_\mathrm{disk}$ --- optical depth in the disk equator at $\lambda=0.5$~$\mu$m, $\beta$ --- exponent in the dependence of disk scale height on distance from the star, $h_0$ --- disk scale height at 10 AU from star, $\epsilon$ --- inclination of the disk.}
  \label{fig:SEDDPVcorner}
\end{figure}

This section presents the posterior distributions of parameters for the models. Fig.~\ref{fig:SEDcorner} contains the results for the model of the spherical envelope obtained using only the SED approximation. Fig.~\ref{fig:SEDDPVcorner} displays the posterior distribution for the parameters of the model of the spherical envelope and disk, derived from SED data and resolved scattered images.

\section{Correction of SED for scattering}
\label{app:SEDcorr}

\begin{figure}[h!]
  \center
  \includegraphics[width=1.0\linewidth]{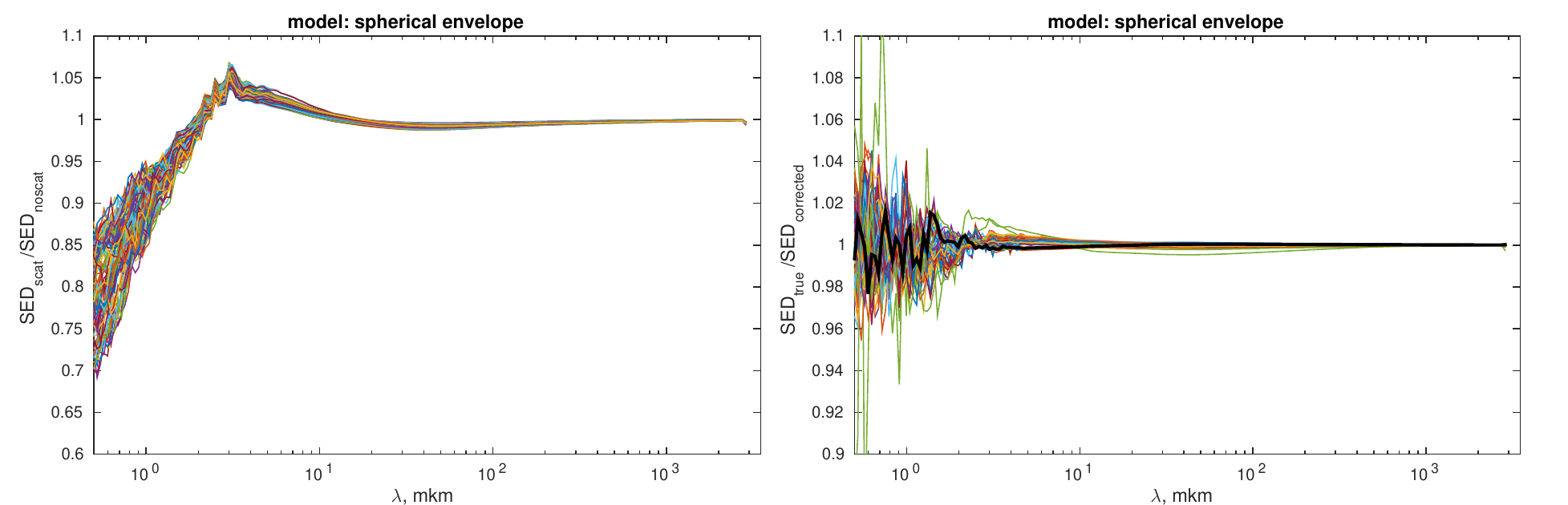}
  \includegraphics[width=1.0\linewidth]{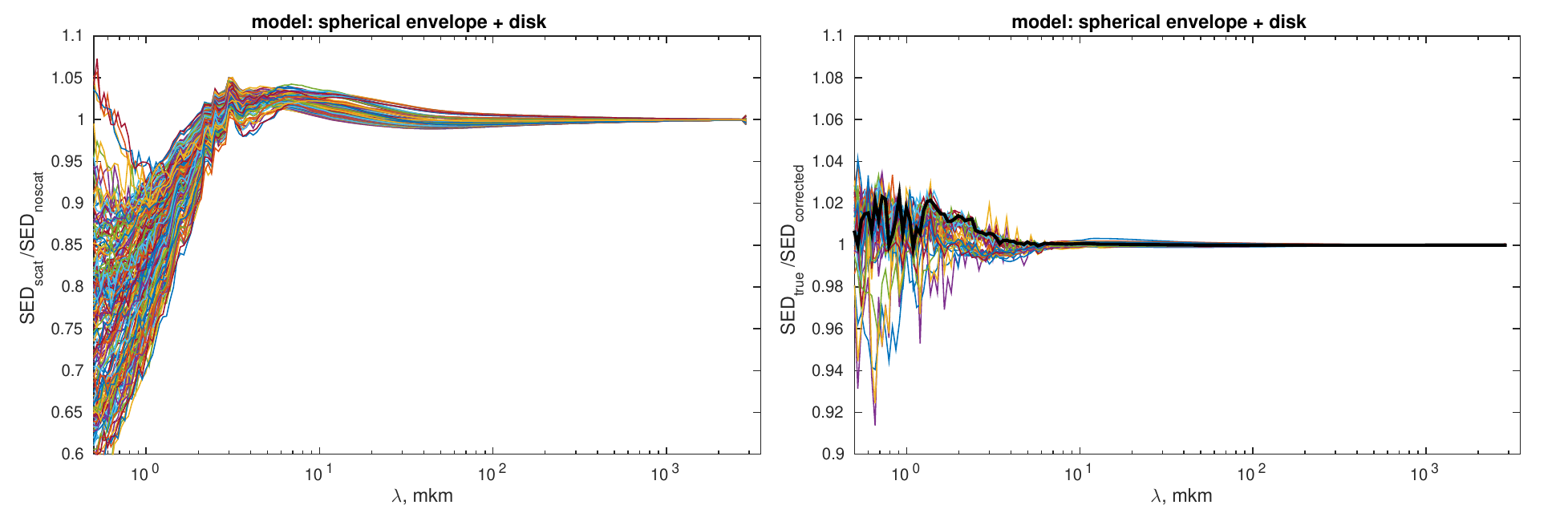}
  \caption{
  Left column: correction factor $\xi=\mathrm{SED}_\mathrm{scat}/\mathrm{SED}_\mathrm{noscat}$ for SED, computed at nodes of the grid presented in Table~\ref{tab:SEDcorr}. Right column: ratios of SED computed using correction method and by direct account for scattering matrices (Section~\ref{subs:SEDapprox}). Thin color lines stand for subsample of volume 50 taken from MCMC generated sample, thick black line stands for optimal model. Upper row: model of spherical envelope, Section~\ref{subs:SEDapprox}, lower row: model of spherical envelope and disk, Section~\ref{subs:DPVapprox}
  }
  \label{fig:SED_correction_sph}
\end{figure}

In sections~\ref{subs:SEDmodel} and \ref{subs:DPVmodel} for the acceleration of the SED computation we employ correction factors $\xi=\mathrm{SED}_\mathrm{scat}/\mathrm{SED}_\mathrm{noscat}$. In table~\ref{tab:SEDcorr} we present the parameters grids for which $\xi$ factors were determined. Fig.~\ref{fig:SED_correction_sph} shows resulting $\xi$ and accuracy of correction.

\begin{deluxetable}{ccc}[h]
\caption{Parameters grids for which factors of the SED correction for the scattering were computed.
\label{tab:SEDcorr}}
\tablewidth{0pt}
\tablehead{\colhead{parameter} & \colhead{spherical envelope} & \colhead{spherical envelope + disk}}
\startdata
$f_\mathrm{C}$           & 0.6667,  0.8571,  0.9091  & 0.7619, 0.8649 \\
$a_\mathrm{max}$,~$\mu$m & 0.5, 0.8, 1.1             & 0.7, 1.1      \\
$b$                      & $-2.1,-2,-1.9$            & $-2$          \\
$r_\mathrm{in}$,~AU      & 10, 14                    & 10            \\
$\tau_\mathrm{sph}$      & 3.5, 4.5                  & 2.5, 4.5 \\
$\tau_\mathrm{disk}$     &                           & 25.0, 45.0    \\
$h_0$,~AU                &                           & 1.2, 2.1      \\
$\beta$                  &                           & $-2.0, -0.8, 0.4$  \\
$\epsilon, ^{\circ}$     &                           & 66, 72        \\
\enddata
\end{deluxetable}

\section{Models with disks at $\beta=0$ and $\beta=1$}
\label{app:disks_beta}

\begin{figure*}
  \center
  \includegraphics[width=0.8\linewidth]{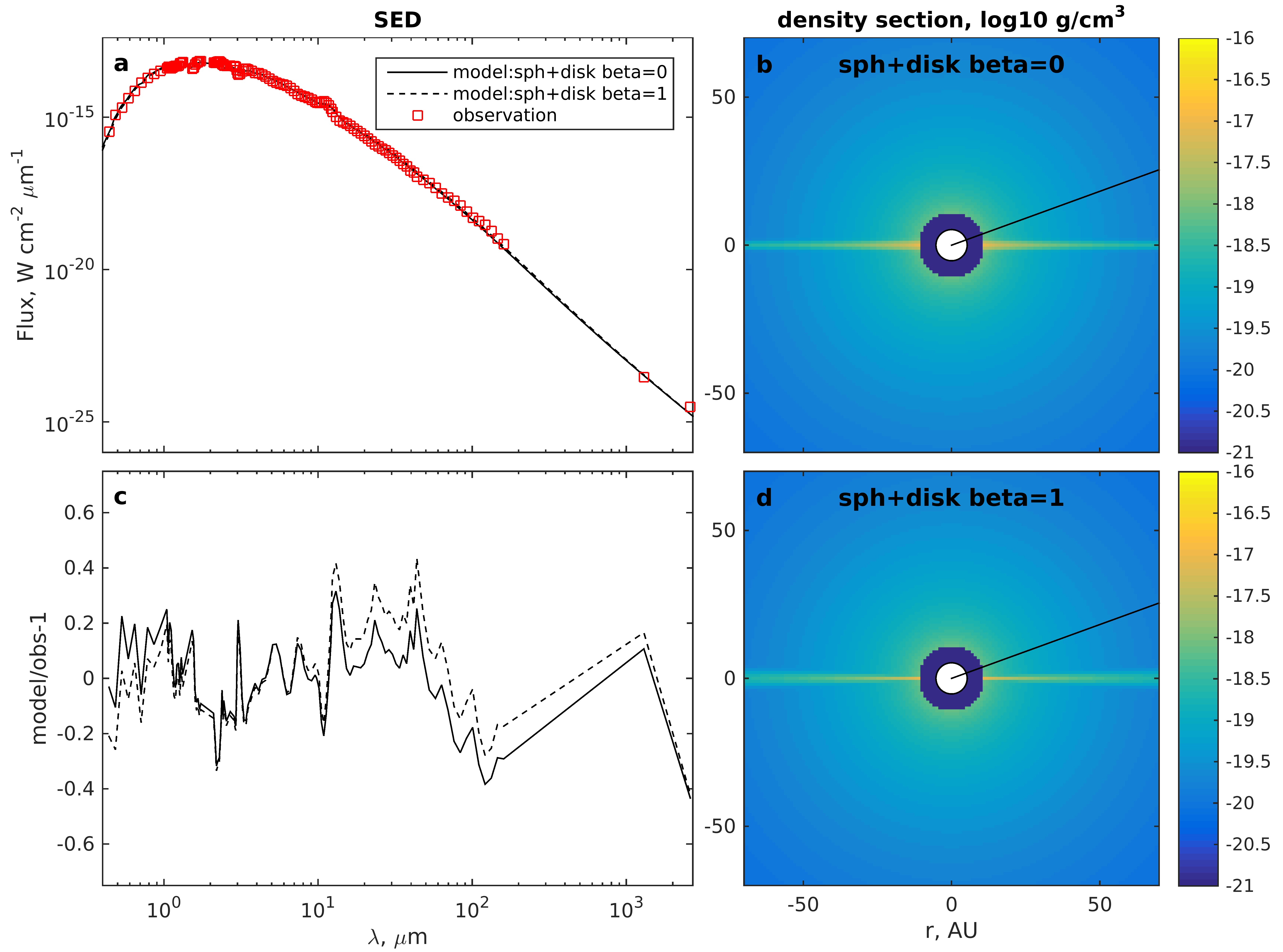}
  \caption{Same as in Fig.~\ref{fig:SED_density}, but for models with disks having $\beta=0$, $h_0=0.7$~AU and $\beta=1$, $h_0=0.2$~AU. 
  \label{fig:SED_density_beta}}
\end{figure*}

\begin{figure*}
  \center
  \includegraphics[width=0.85\linewidth]{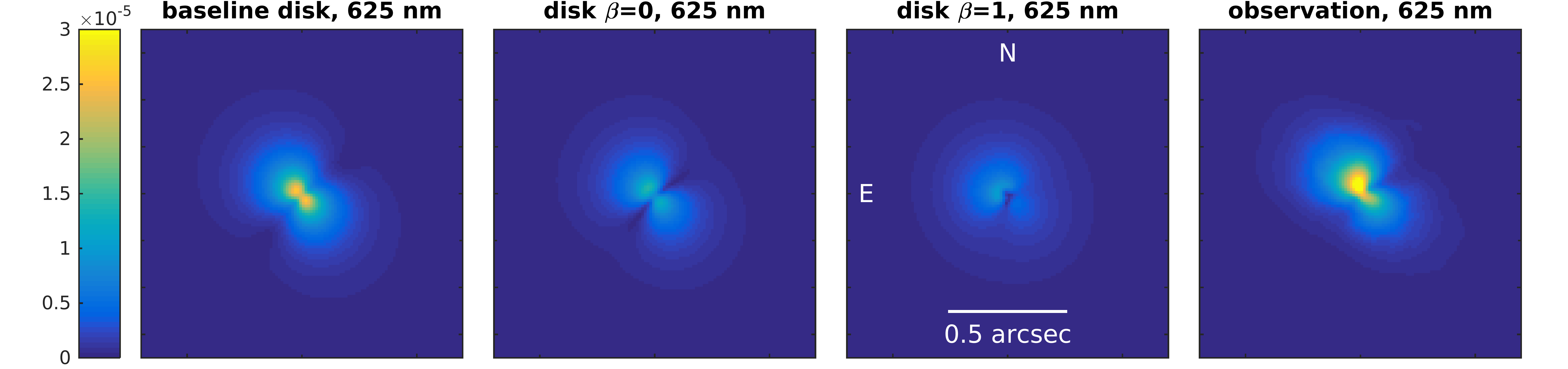}
  \caption{Same as in Fig.~\ref{fig:images_comp}, but for optimal model with disk from Section~\ref{subs:DPVapprox} and for models with disks having $\beta=0$, $h_0=0.7$~AU and $\beta=1$, $h_0=0.2$~AU.}
  \label{fig:images_comp_beta}
\end{figure*}

To test the applicability of models with disks allowing outflow of material we fixed $\beta$ at values $0, 1$. Then for both models we varied the disk scale height $h_0$, searching for maximum posterior probability $p$, as was defined in Sec.~\ref{subs:DPVapprox}. The other parameters of the model was fixed. The resulting optimal values of $h_0$ are 0.7~AU and 0.2~AU for $\beta=0$ and $\beta=1$, respectively. The observables predicted by those models are presented in Fig.~\ref{fig:SED_density_beta} (SED) and \ref{fig:images_comp_beta} (reconstructed images at $\lambda=625$~nm). As can be seen, SEDs are reproduced fairly well, while the images show clear difference.

\newpage

\bibliography{sample631}{}

\begin{thebibliography}{}
\expandafter\ifx\csname natexlab\endcsname\relax\def\natexlab#1{#1}\fi
\providecommand{\url}[1]{\href{#1}{#1}}
\providecommand{\dodoi}[1]{doi:~\href{http://doi.org/#1}{\nolinkurl{#1}}}
\providecommand{\doeprint}[1]{\href{http://ascl.net/#1}{\nolinkurl{http://ascl.net/#1}}}
\providecommand{\doarXiv}[1]{\href{https://arxiv.org/abs/#1}{\nolinkurl{https://arxiv.org/abs/#1}}}

\bibitem[{{Apai} {et~al.}(2004){Apai}, {Pascucci}, {Brandner}, {Henning}, {Lenzen}, {Potter}, {Lagrange}, \& {Rousset}}]{Apai2004}
{Apai}, D., {Pascucci}, I., {Brandner}, W., {et~al.} 2004, \aap, 415, 671, \dodoi{10.1051/0004-6361:20034549}

\bibitem[{{Aringer} {et~al.}(2009){Aringer}, {Girardi}, {Nowotny}, {Marigo}, \& {Lederer}}]{Aringer2009}
{Aringer}, B., {Girardi}, L., {Nowotny}, W., {Marigo}, P., \& {Lederer}, M.~T. 2009, \aap, 503, 913, \dodoi{10.1051/0004-6361/200911703}

\bibitem[{{Astropy Collaboration} {et~al.}(2013){Astropy Collaboration}, {Robitaille}, {Tollerud}, {Greenfield}, {Droettboom}, {Bray}, {Aldcroft}, {Davis}, {Ginsburg}, {Price-Whelan}, {Kerzendorf}, {Conley}, {Crighton}, {Barbary}, {Muna}, {Ferguson}, {Grollier}, {Parikh}, {Nair}, {Unther}, {Deil}, {Woillez}, {Conseil}, {Kramer}, {Turner}, {Singer}, {Fox}, {Weaver}, {Zabalza}, {Edwards}, {Azalee Bostroem}, {Burke}, {Casey}, {Crawford}, {Dencheva}, {Ely}, {Jenness}, {Labrie}, {Lim}, {Pierfederici}, {Pontzen}, {Ptak}, {Refsdal}, {Servillat}, \& {Streicher}}]{2013A&A...558A..33A}
{Astropy Collaboration}, {Robitaille}, T.~P., {Tollerud}, E.~J., {et~al.} 2013, \aap, 558, A33, \dodoi{10.1051/0004-6361/201322068}

\bibitem[{{Astropy Collaboration} {et~al.}(2018){Astropy Collaboration}, {Price-Whelan}, {Sip{\H{o}}cz}, {G{\"u}nther}, {Lim}, {Crawford}, {Conseil}, {Shupe}, {Craig}, {Dencheva}, {Ginsburg}, {VanderPlas}, {Bradley}, {P{\'e}rez-Su{\'a}rez}, {de Val-Borro}, {Aldcroft}, {Cruz}, {Robitaille}, {Tollerud}, {Ardelean}, {Babej}, {Bach}, {Bachetti}, {Bakanov}, {Bamford}, {Barentsen}, {Barmby}, {Baumbach}, {Berry}, {Biscani}, {Boquien}, {Bostroem}, {Bouma}, {Brammer}, {Bray}, {Breytenbach}, {Buddelmeijer}, {Burke}, {Calderone}, {Cano Rodr{\'\i}guez}, {Cara}, {Cardoso}, {Cheedella}, {Copin}, {Corrales}, {Crichton}, {D'Avella}, {Deil}, {Depagne}, {Dietrich}, {Donath}, {Droettboom}, {Earl}, {Erben}, {Fabbro}, {Ferreira}, {Finethy}, {Fox}, {Garrison}, {Gibbons}, {Goldstein}, {Gommers}, {Greco}, {Greenfield}, {Groener}, {Grollier}, {Hagen}, {Hirst}, {Homeier}, {Horton}, {Hosseinzadeh}, {Hu}, {Hunkeler}, {Ivezi{\'c}}, {Jain}, {Jenness}, {Kanarek}, {Kendrew}, {Kern}, {Kerzendorf}, {Khvalko}, {King}, {Kirkby}, {Kulkarni},
  {Kumar}, {Lee}, {Lenz}, {Littlefair}, {Ma}, {Macleod}, {Mastropietro}, {McCully}, {Montagnac}, {Morris}, {Mueller}, {Mumford}, {Muna}, {Murphy}, {Nelson}, {Nguyen}, {Ninan}, {N{\"o}the}, {Ogaz}, {Oh}, {Parejko}, {Parley}, {Pascual}, {Patil}, {Patil}, {Plunkett}, {Prochaska}, {Rastogi}, {Reddy Janga}, {Sabater}, {Sakurikar}, {Seifert}, {Sherbert}, {Sherwood-Taylor}, {Shih}, {Sick}, {Silbiger}, {Singanamalla}, {Singer}, {Sladen}, {Sooley}, {Sornarajah}, {Streicher}, {Teuben}, {Thomas}, {Tremblay}, {Turner}, {Terr{\'o}n}, {van Kerkwijk}, {de la Vega}, {Watkins}, {Weaver}, {Whitmore}, {Woillez}, {Zabalza}, \& {Astropy Contributors}}]{2018AJ....156..123A}
{Astropy Collaboration}, {Price-Whelan}, A.~M., {Sip{\H{o}}cz}, B.~M., {et~al.} 2018, \aj, 156, 123, \dodoi{10.3847/1538-3881/aabc4f}

\bibitem[{{Bergeat} {et~al.}(2001){Bergeat}, {Knapik}, \& {Rutily}}]{Bergeat2001}
{Bergeat}, J., {Knapik}, A., \& {Rutily}, B. 2001, \aap, 369, 178, \dodoi{10.1051/0004-6361:20010106}

\bibitem[{{Bjorkman} \& {Wood}(2001)}]{Bjorkman2001}
{Bjorkman}, J.~E., \& {Wood}, K. 2001, \apj, 554, 615, \dodoi{10.1086/321336}

\bibitem[{{Bujarrabal} {et~al.}(2016){Bujarrabal}, {Castro-Carrizo}, {Alcolea}, {Santander-Garc{\'\i}a}, {van Winckel}, \& {S{\'a}nchez Contreras}}]{Bujarrabal2016}
{Bujarrabal}, V., {Castro-Carrizo}, A., {Alcolea}, J., {et~al.} 2016, \aap, 593, A92, \dodoi{10.1051/0004-6361/201628546}

\bibitem[{{Cardelli} {et~al.}(1989){Cardelli}, {Clayton}, \& {Mathis}}]{ccm89}
{Cardelli}, J.~A., {Clayton}, G.~C., \& {Mathis}, J.~S. 1989, \apj, 345, 245, \dodoi{10.1086/167900}

\bibitem[{{Castro-Carrizo} {et~al.}(2010){Castro-Carrizo}, {Quintana-Lacaci}, {Neri}, {Bujarrabal}, {Sch{\"o}ier}, {Winters}, {Olofsson}, {Lindqvist}, {Alcolea}, {Lucas}, \& {Grewing}}]{CastroCarrizo2010}
{Castro-Carrizo}, A., {Quintana-Lacaci}, G., {Neri}, R., {et~al.} 2010, \aap, 523, A59, \dodoi{10.1051/0004-6361/201014755}

\bibitem[{{Cherchneff}(2006)}]{Cherchneff2006}
{Cherchneff}, I. 2006, \aap, 456, 1001, \dodoi{10.1051/0004-6361:20064827}

\bibitem[{{Clarke}(2010)}]{Clarke2010}
{Clarke}, D. 2010, {Stellar Polarimetry}

\bibitem[{{Cohen}(1979)}]{Cohen1979}
{Cohen}, M. 1979, \mnras, 186, 837, \dodoi{10.1093/mnras/186.4.837}

\bibitem[{{Czekala} {et~al.}(2015){Czekala}, {Andrews}, {Mandel}, {Hogg}, \& {Green}}]{Czekala2015}
{Czekala}, I., {Andrews}, S.~M., {Mandel}, K.~S., {Hogg}, D.~W., \& {Green}, G.~M. 2015, \apj, 812, 128, \dodoi{10.1088/0004-637X/812/2/128}

\bibitem[{{Decin}(2021)}]{Decin2021}
{Decin}, L. 2021, \araa, 59, 337, \dodoi{10.1146/annurev-astro-090120-033712}

\bibitem[{{Dominik} {et~al.}(2021){Dominik}, {Min}, \& {Tazaki}}]{Dominik2021}
{Dominik}, C., {Min}, M., \& {Tazaki}, R. 2021, {OpTool: Command-line driven tool for creating complex dust opacities}, Astrophysics Source Code Library, record ascl:2104.010.
\newblock \doeprint{2104.010}

\bibitem[{{Dullemond} {et~al.}(2012){Dullemond}, {Juhasz}, {Pohl}, {Sereshti}, {Shetty}, {Peters}, {Commercon}, \& {Flock}}]{Dullemond2012}
{Dullemond}, C.~P., {Juhasz}, A., {Pohl}, A., {et~al.} 2012, {RADMC-3D: A multi-purpose radiative transfer tool}, Astrophysics Source Code Library, record ascl:1202.015.
\newblock \doeprint{1202.015}

\bibitem[{{Egan} {et~al.}(2003){Egan}, {Price}, {Kraemer}, {Mizuno}, {Carey}, {Wright}, {Engelke}, {Cohen}, \& {Gugliotti}}]{msx}
{Egan}, M.~P., {Price}, S.~D., {Kraemer}, K.~E., {et~al.} 2003, {VizieR Online Data Catalog: MSX6C Infrared Point Source Catalog. The Midcourse Space Experiment Point Source Catalog Version 2.3 (October 2003)}, VizieR On-line Data Catalog: V/114. Originally published in: Air Force Research Laboratory Technical Report AFRL-VS-TR-2003-1589 (2003)

\bibitem[{{Etma{\'n}ski} {et~al.}(2020){Etma{\'n}ski}, {Schmidt}, \& {Szczerba}}]{Etmanski2020}
{Etma{\'n}ski}, B., {Schmidt}, M.~R., \& {Szczerba}, R. 2020, Advances in Astronomy and Space Physics, 10, 7, \dodoi{10.17721/2227-1481.10.7-11}

\bibitem[{{Fedoteva} {et~al.}(2020){Fedoteva}, {Tatarnikov}, {Safonov}, {Shenavrin}, \& {Komissarova}}]{Fedoteva2020}
{Fedoteva}, A.~A., {Tatarnikov}, A.~M., {Safonov}, B.~S., {Shenavrin}, V.~I., \& {Komissarova}, G.~V. 2020, Astronomy Letters, 46, 38, \dodoi{10.1134/S1063773720010016}

\bibitem[{{Forrest} {et~al.}(1975){Forrest}, {Gillett}, \& {Stein}}]{Forrest1975}
{Forrest}, W.~J., {Gillett}, F.~C., \& {Stein}, W.~A. 1975, \apj, 195, 423, \dodoi{10.1086/153342}

\bibitem[{{Gaia Collaboration} {et~al.}(2016){Gaia Collaboration}, {Prusti}, {de Bruijne}, {Brown}, {Vallenari}, {Babusiaux}, {Bailer-Jones}, {Bastian}, {Biermann}, {Evans}, {Eyer}, {Jansen}, {Jordi}, {Klioner}, {Lammers}, {Lindegren}, {Luri}, {Mignard}, {Milligan}, {Panem}, {Poinsignon}, {Pourbaix}, {Randich}, {Sarri}, {Sartoretti}, {Siddiqui}, {Soubiran}, {Valette}, {van Leeuwen}, {Walton}, {Aerts}, {Arenou}, {Cropper}, {Drimmel}, {H{\o}g}, {Katz}, {Lattanzi}, {O'Mullane}, {Grebel}, {Holland}, {Huc}, {Passot}, {Bramante}, {Cacciari}, {Casta{\~n}eda}, {Chaoul}, {Cheek}, {De Angeli}, {Fabricius}, {Guerra}, {Hern{\'a}ndez}, {Jean-Antoine-Piccolo}, {Masana}, {Messineo}, {Mowlavi}, {Nienartowicz}, {Ord{\'o}{\~n}ez-Blanco}, {Panuzzo}, {Portell}, {Richards}, {Riello}, {Seabroke}, {Tanga}, {Th{\'e}venin}, {Torra}, {Els}, {Gracia-Abril}, {Comoretto}, {Garcia-Reinaldos}, {Lock}, {Mercier}, {Altmann}, {Andrae}, {Astraatmadja}, {Bellas-Velidis}, {Benson}, {Berthier}, {Blomme}, {Busso}, {Carry}, {Cellino}, {Clementini},
  {Cowell}, {Creevey}, {Cuypers}, {Davidson}, {De Ridder}, {de Torres}, {Delchambre}, {Dell'Oro}, {Ducourant}, {Fr{\'e}mat}, {Garc{\'\i}a-Torres}, {Gosset}, {Halbwachs}, {Hambly}, {Harrison}, {Hauser}, {Hestroffer}, {Hodgkin}, {Huckle}, {Hutton}, {Jasniewicz}, {Jordan}, {Kontizas}, {Korn}, {Lanzafame}, {Manteiga}, {Moitinho}, {Muinonen}, {Osinde}, {Pancino}, {Pauwels}, {Petit}, {Recio-Blanco}, {Robin}, {Sarro}, {Siopis}, {Smith}, {Smith}, {Sozzetti}, {Thuillot}, {van Reeven}, {Viala}, {Abbas}, {Abreu Aramburu}, {Accart}, {Aguado}, {Allan}, {Allasia}, {Altavilla}, {{\'A}lvarez}, {Alves}, {Anderson}, {Andrei}, {Anglada Varela}, {Antiche}, {Antoja}, {Ant{\'o}n}, {Arcay}, {Atzei}, {Ayache}, {Bach}, {Baker}, {Balaguer-N{\'u}{\~n}ez}, {Barache}, {Barata}, {Barbier}, {Barblan}, {Baroni}, {Barrado y Navascu{\'e}s}, {Barros}, {Barstow}, {Becciani}, {Bellazzini}, {Bellei}, {Bello Garc{\'\i}a}, {Belokurov}, {Bendjoya}, {Berihuete}, {Bianchi}, {Bienaym{\'e}}, {Billebaud}, {Blagorodnova}, {Blanco-Cuaresma}, {Boch},
  {Bombrun}, {Borrachero}, {Bouquillon}, {Bourda}, {Bouy}, {Bragaglia}, {Breddels}, {Brouillet}, {Br{\"u}semeister}, {Bucciarelli}, {Budnik}, {Burgess}, {Burgon}, {Burlacu}, {Busonero}, {Buzzi}, {Caffau}, {Cambras}, {Campbell}, {Cancelliere}, {Cantat-Gaudin}, {Carlucci}, {Carrasco}, {Castellani}, {Charlot}, {Charnas}, {Charvet}, {Chassat}, {Chiavassa}, {Clotet}, {Cocozza}, {Collins}, {Collins}, {Costigan}, {Crifo}, {Cross}, {Crosta}, {Crowley}, {Dafonte}, {Damerdji}, {Dapergolas}, {David}, {David}, {De Cat}, {de Felice}, {de Laverny}, {De Luise}, {De March}, {de Martino}, {de Souza}, {Debosscher}, {del Pozo}, {Delbo}, {Delgado}, {Delgado}, {di Marco}, {Di Matteo}, {Diakite}, {Distefano}, {Dolding}, {Dos Anjos}, {Drazinos}, {Dur{\'a}n}, {Dzigan}, {Ecale}, {Edvardsson}, {Enke}, {Erdmann}, {Escolar}, {Espina}, {Evans}, {Eynard Bontemps}, {Fabre}, {Fabrizio}, {Faigler}, {Falc{\~a}o}, {Farr{\`a}s Casas}, {Faye}, {Federici}, {Fedorets}, {Fern{\'a}ndez-Hern{\'a}ndez}, {Fernique}, {Fienga}, {Figueras}, {Filippi},
  {Findeisen}, {Fonti}, {Fouesneau}, {Fraile}, {Fraser}, {Fuchs}, {Furnell}, {Gai}, {Galleti}, {Galluccio}, {Garabato}, {Garc{\'\i}a-Sedano}, {Gar{\'e}}, {Garofalo}, {Garralda}, {Gavras}, {Gerssen}, {Geyer}, {Gilmore}, {Girona}, {Giuffrida}, {Gomes}, {Gonz{\'a}lez-Marcos}, {Gonz{\'a}lez-N{\'u}{\~n}ez}, {Gonz{\'a}lez-Vidal}, {Granvik}, {Guerrier}, {Guillout}, {Guiraud}, {G{\'u}rpide}, {Guti{\'e}rrez-S{\'a}nchez}, {Guy}, {Haigron}, {Hatzidimitriou}, {Haywood}, {Heiter}, {Helmi}, {Hobbs}, {Hofmann}, {Holl}, {Holland}, {Hunt}, {Hypki}, {Icardi}, {Irwin}, {Jevardat de Fombelle}, {Jofr{\'e}}, {Jonker}, {Jorissen}, {Julbe}, {Karampelas}, {Kochoska}, {Kohley}, {Kolenberg}, {Kontizas}, {Koposov}, {Kordopatis}, {Koubsky}, {Kowalczyk}, {Krone-Martins}, {Kudryashova}, {Kull}, {Bachchan}, {Lacoste-Seris}, {Lanza}, {Lavigne}, {Le Poncin-Lafitte}, {Lebreton}, {Lebzelter}, {Leccia}, {Leclerc}, {Lecoeur-Taibi}, {Lemaitre}, {Lenhardt}, {Leroux}, {Liao}, {Licata}, {Lindstr{\o}m}, {Lister}, {Livanou}, {Lobel}, {L{\"o}ffler},
  {L{\'o}pez}, {Lopez-Lozano}, {Lorenz}, {Loureiro}, {MacDonald}, {Magalh{\~a}es Fernandes}, {Managau}, {Mann}, {Mantelet}, {Marchal}, {Marchant}, {Marconi}, {Marie}, {Marinoni}, {Marrese}, {Marschalk{\'o}}, {Marshall}, {Mart{\'\i}n-Fleitas}, {Martino}, {Mary}, {Matijevi{\v{c}}}, {Mazeh}, {McMillan}, {Messina}, {Mestre}, {Michalik}, {Millar}, {Miranda}, {Molina}, {Molinaro}, {Molinaro}, {Moln{\'a}r}, {Moniez}, {Montegriffo}, {Monteiro}, {Mor}, {Mora}, {Morbidelli}, {Morel}, {Morgenthaler}, {Morley}, {Morris}, {Mulone}, {Muraveva}, {Musella}, {Narbonne}, {Nelemans}, {Nicastro}, {Noval}, {Ord{\'e}novic}, {Ordieres-Mer{\'e}}, {Osborne}, {Pagani}, {Pagano}, {Pailler}, {Palacin}, {Palaversa}, {Parsons}, {Paulsen}, {Pecoraro}, {Pedrosa}, {Pentik{\"a}inen}, {Pereira}, {Pichon}, {Piersimoni}, {Pineau}, {Plachy}, {Plum}, {Poujoulet}, {Pr{\v{s}}a}, {Pulone}, {Ragaini}, {Rago}, {Rambaux}, {Ramos-Lerate}, {Ranalli}, {Rauw}, {Read}, {Regibo}, {Renk}, {Reyl{\'e}}, {Ribeiro}, {Rimoldini}, {Ripepi}, {Riva}, {Rixon},
  {Roelens}, {Romero-G{\'o}mez}, {Rowell}, {Royer}, {Rudolph}, {Ruiz-Dern}, {Sadowski}, {Sagrist{\`a} Sell{\'e}s}, {Sahlmann}, {Salgado}, {Salguero}, {Sarasso}, {Savietto}, {Schnorhk}, {Schultheis}, {Sciacca}, {Segol}, {Segovia}, {Segransan}, {Serpell}, {Shih}, {Smareglia}, {Smart}, {Smith}, {Solano}, {Solitro}, {Sordo}, {Soria Nieto}, {Souchay}, {Spagna}, {Spoto}, {Stampa}, {Steele}, {Steidelm{\"u}ller}, {Stephenson}, {Stoev}, {Suess}, {S{\"u}veges}, {Surdej}, {Szabados}, {Szegedi-Elek}, {Tapiador}, {Taris}, {Tauran}, {Taylor}, {Teixeira}, {Terrett}, {Tingley}, {Trager}, {Turon}, {Ulla}, {Utrilla}, {Valentini}, {van Elteren}, {Van Hemelryck}, {van Leeuwen}, {Varadi}, {Vecchiato}, {Veljanoski}, {Via}, {Vicente}, {Vogt}, {Voss}, {Votruba}, {Voutsinas}, {Walmsley}, {Weiler}, {Weingrill}, {Werner}, {Wevers}, {Whitehead}, {Wyrzykowski}, {Yoldas}, {{\v{Z}}erjal}, {Zucker}, {Zurbach}, {Zwitter}, {Alecu}, {Allen}, {Allende Prieto}, {Amorim}, {Anglada-Escud{\'e}}, {Arsenijevic}, {Azaz}, {Balm}, {Beck}, {Bernstein},
  {Bigot}, {Bijaoui}, {Blasco}, {Bonfigli}, {Bono}, {Boudreault}, {Bressan}, {Brown}, {Brunet}, {Bunclark}, {Buonanno}, {Butkevich}, {Carret}, {Carrion}, {Chemin}, {Ch{\'e}reau}, {Corcione}, {Darmigny}, {de Boer}, {de Teodoro}, {de Zeeuw}, {Delle Luche}, {Domingues}, {Dubath}, {Fodor}, {Fr{\'e}zouls}, {Fries}, {Fustes}, {Fyfe}, {Gallardo}, {Gallegos}, {Gardiol}, {Gebran}, {Gomboc}, {G{\'o}mez}, {Grux}, {Gueguen}, {Heyrovsky}, {Hoar}, {Iannicola}, {Isasi Parache}, {Janotto}, {Joliet}, {Jonckheere}, {Keil}, {Kim}, {Klagyivik}, {Klar}, {Knude}, {Kochukhov}, {Kolka}, {Kos}, {Kutka}, {Lainey}, {LeBouquin}, {Liu}, {Loreggia}, {Makarov}, {Marseille}, {Martayan}, {Martinez-Rubi}, {Massart}, {Meynadier}, {Mignot}, {Munari}, {Nguyen}, {Nordlander}, {Ocvirk}, {O'Flaherty}, {Olias Sanz}, {Ortiz}, {Osorio}, {Oszkiewicz}, {Ouzounis}, {Palmer}, {Park}, {Pasquato}, {Peltzer}, {Peralta}, {P{\'e}turaud}, {Pieniluoma}, {Pigozzi}, {Poels}, {Prat}, {Prod'homme}, {Raison}, {Rebordao}, {Risquez}, {Rocca-Volmerange}, {Rosen},
  {Ruiz-Fuertes}, {Russo}, {Sembay}, {Serraller Vizcaino}, {Short}, {Siebert}, {Silva}, {Sinachopoulos}, {Slezak}, {Soffel}, {Sosnowska}, {Strai{\v{z}}ys}, {ter Linden}, {Terrell}, {Theil}, {Tiede}, {Troisi}, {Tsalmantza}, {Tur}, {Vaccari}, {Vachier}, {Valles}, {Van Hamme}, {Veltz}, {Virtanen}, {Wallut}, {Wichmann}, {Wilkinson}, {Ziaeepour}, \& {Zschocke}}]{gaia1}
{Gaia Collaboration}, {Prusti}, T., {de Bruijne}, J.~H.~J., {et~al.} 2016, \aap, 595, A1, \dodoi{10.1051/0004-6361/201629272}

\bibitem[{{Gaia Collaboration} {et~al.}(2022){Gaia Collaboration}, {Vallenari}, {Brown}, {Prusti}, {de Bruijne}, {Arenou}, {Babusiaux}, {Biermann}, {Creevey}, {Ducourant}, {Evans}, {Eyer}, {Guerra}, {Hutton}, {Jordi}, {Klioner}, {Lammers}, {Lindegren}, {Luri}, {Mignard}, {Panem}, {Pourbaix}, {Randich}, {Sartoretti}, {Soubiran}, {Tanga}, {Walton}, {Bailer-Jones}, {Bastian}, {Drimmel}, {Jansen}, {Katz}, {Lattanzi}, {van Leeuwen}, {Bakker}, {Cacciari}, {Casta{\~n}eda}, {De Angeli}, {Fabricius}, {Fouesneau}, {Fr{\'e}mat}, {Galluccio}, {Guerrier}, {Heiter}, {Masana}, {Messineo}, {Mowlavi}, {Nicolas}, {Nienartowicz}, {Pailler}, {Panuzzo}, {Riclet}, {Roux}, {Seabroke}, {Sordo{\o}rcit}, {Th{\'e}venin}, {Gracia-Abril}, {Portell}, {Teyssier}, {Altmann}, {Andrae}, {Audard}, {Bellas-Velidis}, {Benson}, {Berthier}, {Blomme}, {Burgess}, {Busonero}, {Busso}, {C{\'a}novas}, {Carry}, {Cellino}, {Cheek}, {Clementini}, {Damerdji}, {Davidson}, {de Teodoro}, {Nu{\~n}ez Campos}, {Delchambre}, {Dell'Oro}, {Esquej},
  {Fern{\'a}ndez-Hern{\'a}ndez}, {Fraile}, {Garabato}, {Garc{\'\i}a-Lario}, {Gosset}, {Haigron}, {Halbwachs}, {Hambly}, {Harrison}, {Hern{\'a}ndez}, {Hestroffer}, {Hodgkin}, {Holl}, {Jan{\ss}en}, {Jevardat de Fombelle}, {Jordan}, {Krone-Martins}, {Lanzafame}, {L{\"o}ffler}, {Marchal}, {Marrese}, {Moitinho}, {Muinonen}, {Osborne}, {Pancino}, {Pauwels}, {Recio-Blanco}, {Reyl{\'e}}, {Riello}, {Rimoldini}, {Roegiers}, {Rybizki}, {Sarro}, {Siopis}, {Smith}, {Sozzetti}, {Utrilla}, {van Leeuwen}, {Abbas}, {{\'A}brah{\'a}m}, {Abreu Aramburu}, {Aerts}, {Aguado}, {Ajaj}, {Aldea-Montero}, {Altavilla}, {{\'A}lvarez}, {Alves}, {Anders}, {Anderson}, {Anglada Varela}, {Antoja}, {Baines}, {Baker}, {Balaguer-N{\'u}{\~n}ez}, {Balbinot}, {Balog}, {Barache}, {Barbato}, {Barros}, {Barstow}, {Bartolom{\'e}}, {Bassilana}, {Bauchet}, {Becciani}, {Bellazzini}, {Berihuete}, {Bernet}, {Bertone}, {Bianchi}, {Binnenfeld}, {Blanco-Cuaresma}, {Blazere}, {Boch}, {Bombrun}, {Bossini}, {Bouquillon}, {Bragaglia}, {Bramante}, {Breedt},
  {Bressan}, {Brouillet}, {Brugaletta}, {Bucciarelli}, {Burlacu}, {Butkevich}, {Buzzi}, {Caffau}, {Cancelliere}, {Cantat-Gaudin}, {Carballo}, {Carlucci}, {Carnerero}, {Carrasco}, {Casamiquela}, {Castellani}, {Castro-Ginard}, {Chaoul}, {Charlot}, {Chemin}, {Chiaramida}, {Chiavassa}, {Chornay}, {Comoretto}, {Contursi}, {Cooper}, {Cornez}, {Cowell}, {Crifo}, {Cropper}, {Crosta}, {Crowley}, {Dafonte}, {Dapergolas}, {David}, {David}, {de Laverny}, {De Luise}, {De March}, {De Ridder}, {de Souza}, {de Torres}, {del Peloso}, {del Pozo}, {Delbo}, {Delgado}, {Delisle}, {Demouchy}, {Dharmawardena}, {Di Matteo}, {Diakite}, {Diener}, {Distefano}, {Dolding}, {Edvardsson}, {Enke}, {Fabre}, {Fabrizio}, {Faigler}, {Fedorets}, {Fernique}, {Fienga}, {Figueras}, {Fournier}, {Fouron}, {Fragkoudi}, {Gai}, {Garcia-Gutierrez}, {Garcia-Reinaldos}, {Garc{\'\i}a-Torres}, {Garofalo}, {Gavel}, {Gavras}, {Gerlach}, {Geyer}, {Giacobbe}, {Gilmore}, {Girona}, {Giuffrida}, {Gomel}, {Gomez}, {Gonz{\'a}lez-N{\'u}{\~n}ez},
  {Gonz{\'a}lez-Santamar{\'\i}a}, {Gonz{\'a}lez-Vidal}, {Granvik}, {Guillout}, {Guiraud}, {Guti{\'e}rrez-S{\'a}nchez}, {Guy}, {Hatzidimitriou}, {Hauser}, {Haywood}, {Helmer}, {Helmi}, {Sarmiento}, {Hidalgo}, {Hilger}, {H{\l}adczuk}, {Hobbs}, {Holland}, {Huckle}, {Jardine}, {Jasniewicz}, {Jean-Antoine Piccolo}, {Jim{\'e}nez-Arranz}, {Jorissen}, {Juaristi Campillo}, {Julbe}, {Karbevska}, {Kervella}, {Khanna}, {Kontizas}, {Kordopatis}, {Korn}, {K{\'o}sp{\'a}l}, {Kostrzewa-Rutkowska}, {Kruszy{\'n}ska}, {Kun}, {Laizeau}, {Lambert}, {Lanza}, {Lasne}, {Le Campion}, {Lebreton}, {Lebzelter}, {Leccia}, {Leclerc}, {Lecoeur-Taibi}, {Liao}, {Licata}, {Lindstr{\o}m}, {Lister}, {Livanou}, {Lobel}, {Lorca}, {Loup}, {Madrero Pardo}, {Magdaleno Romeo}, {Managau}, {Mann}, {Manteiga}, {Marchant}, {Marconi}, {Marcos}, {Marcos Santos}, {Mar{\'\i}n Pina}, {Marinoni}, {Marocco}, {Marshall}, {Polo}, {Mart{\'\i}n-Fleitas}, {Marton}, {Mary}, {Masip}, {Massari}, {Mastrobuono-Battisti}, {Mazeh}, {McMillan}, {Messina}, {Michalik},
  {Millar}, {Mints}, {Molina}, {Molinaro}, {Moln{\'a}r}, {Monari}, {Mongui{\'o}}, {Montegriffo}, {Montero}, {Mor}, {Mora}, {Morbidelli}, {Morel}, {Morris}, {Muraveva}, {Murphy}, {Musella}, {Nagy}, {Noval}, {Oca{\~n}a}, {Ogden}, {Ordenovic}, {Osinde}, {Pagani}, {Pagano}, {Palaversa}, {Palicio}, {Pallas-Quintela}, {Panahi}, {Payne-Wardenaar}, {Pe{\~n}alosa Esteller}, {Penttil{\"a}}, {Pichon}, {Piersimoni}, {Pineau}, {Plachy}, {Plum}, {Poggio}, {Pr{\v{s}}a}, {Pulone}, {Racero}, {Ragaini}, {Rainer}, {Raiteri}, {Rambaux}, {Ramos}, {Ramos-Lerate}, {Re Fiorentin}, {Regibo}, {Richards}, {Rios Diaz}, {Ripepi}, {Riva}, {Rix}, {Rixon}, {Robichon}, {Robin}, {Robin}, {Roelens}, {Rogues}, {Rohrbasser}, {Romero-G{\'o}mez}, {Rowell}, {Royer}, {Ruz Mieres}, {Rybicki}, {Sadowski}, {S{\'a}ez N{\'u}{\~n}ez}, {Sagrist{\`a} Sell{\'e}s}, {Sahlmann}, {Salguero}, {Samaras}, {Sanchez Gimenez}, {Sanna}, {Santove{\~n}a}, {Sarasso}, {Schultheis}, {Sciacca}, {Segol}, {Segovia}, {S{\'e}gransan}, {Semeux}, {Shahaf}, {Siddiqui}, {Siebert},
  {Siltala}, {Silvelo}, {Slezak}, {Slezak}, {Smart}, {Snaith}, {Solano}, {Solitro}, {Souami}, {Souchay}, {Spagna}, {Spina}, {Spoto}, {Steele}, {Steidelm{\"u}ller}, {Stephenson}, {S{\"u}veges}, {Surdej}, {Szabados}, {Szegedi-Elek}, {Taris}, {Taylo}, {Teixeira}, {Tolomei}, {Tonello}, {Torra}, {Torra}, {Torralba Elipe}, {Trabucchi}, {Tsounis}, {Turon}, {Ulla}, {Unger}, {Vaillant}, {van Dillen}, {van Reeven}, {Vanel}, {Vecchiato}, {Viala}, {Vicente}, {Voutsinas}, {Weiler}, {Wevers}, {Wyrzykowski}, {Yoldas}, {Yvard}, {Zhao}, {Zorec}, {Zucker}, \& {Zwitter}}]{gaiadr3}
{Gaia Collaboration}, {Vallenari}, A., {Brown}, A.~G.~A., {et~al.} 2022, arXiv e-prints, arXiv:2208.00211.
\newblock \doarXiv{2208.00211}

\bibitem[{{Garufi} {et~al.}(2024){Garufi}, {Ginski}, {van Holstein}, {Benisty}, {Manara}, {P{\'e}rez}, {Pinilla}, {Ribas}, {Weber}, {Williams}, {Cieza}, {Dominik}, {Facchini}, {Huang}, {Zurlo}, {Bae}, {Hagelberg}, {Henning}, {Hogerheijde}, {Janson}, {M{\'e}nard}, {Messina}, {Meyer}, {Pinte}, {Quanz}, {Rigliaco}, {Roccatagliata}, {Schmid}, {Szul{\'a}gyi}, {van Boekel}, {Wahhaj}, {Antichi}, {Baruffolo}, \& {Moulin}}]{Garufi2024}
{Garufi}, A., {Ginski}, C., {van Holstein}, R.~G., {et~al.} 2024, \aap, 685, A53, \dodoi{10.1051/0004-6361/202347586}

\bibitem[{{Green} {et~al.}(2019){Green}, {Schlafly}, {Zucker}, {Speagle}, \& {Finkbeiner}}]{dustmap}
{Green}, G.~M., {Schlafly}, E., {Zucker}, C., {Speagle}, J.~S., \& {Finkbeiner}, D. 2019, \apj, 887, 93, \dodoi{10.3847/1538-4357/ab5362}

\bibitem[{{Groenewegen} {et~al.}(1998){Groenewegen}, {Whitelock}, {Smith}, \& {Kerschbaum}}]{Groenewegen1998}
{Groenewegen}, M.~A.~T., {Whitelock}, P.~A., {Smith}, C.~H., \& {Kerschbaum}, F. 1998, \mnras, 293, 18, \dodoi{10.1046/j.1365-8711.1998.01113.x}

\bibitem[{{Hackwell}(1972)}]{Hackwell1972}
{Hackwell}, J.~A. 1972, \aap, 21, 239

\bibitem[{{H{\"o}fner} \& {Olofsson}(2018)}]{Hofner2018}
{H{\"o}fner}, S., \& {Olofsson}, H. 2018, \aapr, 26, 1, \dodoi{10.1007/s00159-017-0106-5}

\bibitem[{{H{\o}g} {et~al.}(2000){H{\o}g}, {Fabricius}, {Makarov}, {Urban}, {Corbin}, {Wycoff}, {Bastian}, {Schwekendiek}, \& {Wicenec}}]{tycho2}
{H{\o}g}, E., {Fabricius}, C., {Makarov}, V.~V., {et~al.} 2000, \aap, 355, L27

\bibitem[{{Homan} {et~al.}(2020){Homan}, {Cannon}, {Montarg{\`e}s}, {Richards}, {Millar}, \& {Decin}}]{Homan2020}
{Homan}, W., {Cannon}, E., {Montarg{\`e}s}, M., {et~al.} 2020, \aap, 642, A93, \dodoi{10.1051/0004-6361/202038255}

\bibitem[{{Iben} \& {Renzini}(1983)}]{Iben1983}
{Iben}, I., J., \& {Renzini}, A. 1983, \araa, 21, 271, \dodoi{10.1146/annurev.aa.21.090183.001415}

\bibitem[{{Ishihara} {et~al.}(2010){Ishihara}, {Onaka}, {Kataza}, {Salama}, {Alfageme}, {Cassatella}, {Cox}, {Garc{\'\i}a-Lario}, {Stephenson}, {Cohen}, {Fujishiro}, {Fujiwara}, {Hasegawa}, {Ita}, {Kim}, {Matsuhara}, {Murakami}, {M{\"u}ller}, {Nakagawa}, {Ohyama}, {Oyabu}, {Pyo}, {Sakon}, {Shibai}, {Takita}, {Tanab{\'e}}, {Uemizu}, {Ueno}, {Usui}, {Wada}, {Watarai}, {Yamamura}, \& {Yamauchi}}]{akari}
{Ishihara}, D., {Onaka}, T., {Kataza}, H., {et~al.} 2010, \aap, 514, A1, \dodoi{10.1051/0004-6361/200913811}

\bibitem[{{Kerschbaum} {et~al.}(2010){Kerschbaum}, {Lebzelter}, \& {Mekul}}]{Kerschbaum2010}
{Kerschbaum}, F., {Lebzelter}, T., \& {Mekul}, L. 2010, \aap, 524, A87, \dodoi{10.1051/0004-6361/201014514}

\bibitem[{{Kervella} {et~al.}(2016){Kervella}, {Homan}, {Richards}, {Decin}, {McDonald}, {Montarg{\`e}s}, \& {Ohnaka}}]{Kervella2016}
{Kervella}, P., {Homan}, W., {Richards}, A.~M.~S., {et~al.} 2016, \aap, 596, A92, \dodoi{10.1051/0004-6361/201629877}

\bibitem[{{Kornilov} {et~al.}(2016){Kornilov}, {Kornilov}, {Voziakova}, {Shatsky}, {Safonov}, {Gorbunov}, {Potanin}, {Cheryasov}, \& {Senik}}]{Kornilov2016}
{Kornilov}, V., {Kornilov}, M., {Voziakova}, O., {et~al.} 2016, \mnras, 462, 4464, \dodoi{10.1093/mnras/stw1839}

\bibitem[{{Kuhn} {et~al.}(2001){Kuhn}, {Potter}, \& {Parise}}]{Kuhn2001}
{Kuhn}, J.~R., {Potter}, D., \& {Parise}, B. 2001, \apjl, 553, L189, \dodoi{10.1086/320686}

\bibitem[{{Lombaert} {et~al.}(2016){Lombaert}, {Decin}, {Royer}, {de Koter}, {Cox}, {Gonz{\'a}lez-Alfonso}, {Neufeld}, {De Ridder}, {Ag{\'u}ndez}, {Blommaert}, {Khouri}, {Groenewegen}, {Kerschbaum}, {Cernicharo}, {Vandenbussche}, \& {Waelkens}}]{Lombaert2016}
{Lombaert}, R., {Decin}, L., {Royer}, P., {et~al.} 2016, \aap, 588, A124, \dodoi{10.1051/0004-6361/201527049}

\bibitem[{{Matesic} {et~al.}(2024){Matesic}, {Rowe}, {Livingston}, {Dholakia}, {Jontof-Hutter}, \& {Lissauer}}]{Matesic2024}
{Matesic}, M. R.~B., {Rowe}, J.~F., {Livingston}, J.~H., {et~al.} 2024, \aj, 167, 68, \dodoi{10.3847/1538-3881/ad0fe9}

\bibitem[{{Melikian}(1996)}]{Melikian1996}
{Melikian}, N.~D. 1996, Astrophysics, 39, 321, \dodoi{10.1007/BF02077202}

\bibitem[{{Murakawa} {et~al.}(2010){Murakawa}, {Ueta}, \& {Meixner}}]{Murakawa2010}
{Murakawa}, K., {Ueta}, T., \& {Meixner}, M. 2010, \aap, 510, A30, \dodoi{10.1051/0004-6361/200912674}

\bibitem[{{Nadjip} {et~al.}(2017){Nadjip}, {Tatarnikov}, {Toomey}, {Shatsky}, {Cherepashchuk}, {Lamzin}, \& {Belinski}}]{Nadjip2017}
{Nadjip}, A.~E., {Tatarnikov}, A.~M., {Toomey}, D.~W., {et~al.} 2017, Astrophysical Bulletin, 72, 349, \dodoi{10.1134/S1990341317030245}

\bibitem[{{Neufeld} {et~al.}(2010){Neufeld}, {Gonz{\'a}lez-Alfonso}, {Melnick}, {Pu{\l}ecka}, {Schmidt}, {Szczerba}, {Bujarrabal}, {Alcolea}, {Cernicharo}, {Decin}, {Dominik}, {Justtanont}, {de Koter}, {Marston}, {Menten}, {Olofsson}, {Planesas}, {Sch{\"o}ier}, {Teyssier}, {Waters}, {Edwards}, {McCoey}, {Shipman}, {Jellema}, {de Graauw}, {Ossenkopf}, {Schieder}, \& {Philipp}}]{Neufeld2010}
{Neufeld}, D.~A., {Gonz{\'a}lez-Alfonso}, E., {Melnick}, G., {et~al.} 2010, \aap, 521, L5, \dodoi{10.1051/0004-6361/201015080}

\bibitem[{{Neufeld} {et~al.}(2011){Neufeld}, {Gonz{\'a}lez-Alfonso}, {Melnick}, {Szczerba}, {Schmidt}, {Decin}, {Alcolea}, {de Koter}, {Sch{\"o}ier}, {Bujarrabal}, {Cernicharo}, {Dominik}, {Justtanont}, {Marston}, {Menten}, {Olofsson}, {Planesas}, {Teyssier}, \& {Waters}}]{Neufeld2011}
---. 2011, \apjl, 727, L29, \dodoi{10.1088/2041-8205/727/2/L29}

\bibitem[{{Neugebauer} {et~al.}(1984){Neugebauer}, {Habing}, {van Duinen}, {Aumann}, {Baud}, {Beichman}, {Beintema}, {Boggess}, {Clegg}, {de Jong}, {Emerson}, {Gautier}, {Gillett}, {Harris}, {Hauser}, {Houck}, {Jennings}, {Low}, {Marsden}, {Miley}, {Olnon}, {Pottasch}, {Raimond}, {Rowan-Robinson}, {Soifer}, {Walker}, {Wesselius}, \& {Young}}]{iras}
{Neugebauer}, G., {Habing}, H.~J., {van Duinen}, R., {et~al.} 1984, \apjl, 278, L1, \dodoi{10.1086/184209}

\bibitem[{{Nhung} {et~al.}(2024){Nhung}, {Hoai}, {Darriulat}, {Tuan-Anh}, {Diep}, {Ngoc}, \& {Thai}}]{Nhung2024}
{Nhung}, P.~T., {Hoai}, D.~T., {Darriulat}, P., {et~al.} 2024, Research in Astronomy and Astrophysics, 24, 035009, \dodoi{10.1088/1674-4527/ad18a4}

\bibitem[{{Norris} {et~al.}(2012){Norris}, {Tuthill}, {Ireland}, {Lacour}, {Zijlstra}, {Lykou}, {Evans}, {Stewart}, \& {Bedding}}]{Norris2012}
{Norris}, B. R.~M., {Tuthill}, P.~G., {Ireland}, M.~J., {et~al.} 2012, \nat, 484, 220, \dodoi{10.1038/nature10935}

\bibitem[{{Oppenheimer} {et~al.}(2005){Oppenheimer}, {Bieging}, {Schmidt}, {Gordon}, {Misselt}, \& {Smith}}]{Oppenheimer2005}
{Oppenheimer}, B.~D., {Bieging}, J.~H., {Schmidt}, G.~D., {et~al.} 2005, \apj, 624, 957, \dodoi{10.1086/429081}

\bibitem[{{Pegourie}(1988)}]{Pegourie1988}
{Pegourie}, B. 1988, \aap, 194, 335

\bibitem[{{Safonov} {et~al.}(2019{\natexlab{a}}){Safonov}, {Lysenko}, {Goliguzova}, \& {Cheryasov}}]{Safonov2019}
{Safonov}, B., {Lysenko}, P., {Goliguzova}, M., \& {Cheryasov}, D. 2019{\natexlab{a}}, \mnras, 484, 5129, \dodoi{10.1093/mnras/stz288}

\bibitem[{{Safonov} {et~al.}(2023){Safonov}, {Millar-Blanchaer}, {Zhang}, {Norris}, {Guyon}, {Lozi}, \& {Sallum}}]{Safonov2023}
{Safonov}, B., {Millar-Blanchaer}, M.~A., {Zhang}, M., {et~al.} 2023, Journal of Astronomical Telescopes, Instruments, and Systems, 9, 028005, \dodoi{10.1117/1.JATIS.9.2.028005}

\bibitem[{{Safonov} {et~al.}(2020){Safonov}, {Dodin}, {Burlak}, {Goliguzova}, {Fedoteva}, {Zheltoukhov}, {Lamzin}, {Strakhov}, \& {Voziakova}}]{Safonov2020}
{Safonov}, B., {Dodin}, A., {Burlak}, M., {et~al.} 2020, arXiv e-prints, arXiv:2005.05215, \dodoi{10.48550/arXiv.2005.05215}

\bibitem[{{Safonov} {et~al.}(2019{\natexlab{b}}){Safonov}, {Dodin}, {Lamzin}, \& {Rastorguev}}]{Safonov2019b}
{Safonov}, B.~S., {Dodin}, A.~V., {Lamzin}, S.~A., \& {Rastorguev}, A.~S. 2019{\natexlab{b}}, Astronomy Letters, 45, 453, \dodoi{10.1134/S1063773719070065}

\bibitem[{{Safonov} {et~al.}(2017){Safonov}, {Lysenko}, \& {Dodin}}]{Safonov2017}
{Safonov}, B.~S., {Lysenko}, P.~A., \& {Dodin}, A.~V. 2017, Astronomy Letters, 43, 344, \dodoi{10.1134/S1063773717050036}

\bibitem[{{Safonov} {et~al.}(2022){Safonov}, {Strakhov}, {Goliguzova}, \& {Voziakova}}]{Safonov2022}
{Safonov}, B.~S., {Strakhov}, I.~A., {Goliguzova}, M.~V., \& {Voziakova}, O.~V. 2022, \aj, 163, 31, \dodoi{10.3847/1538-3881/ac36cb}

\bibitem[{{Samus'} {et~al.}(2017){Samus'}, {Kazarovets}, {Durlevich}, {Kireeva}, \& {Pastukhova}}]{Samus2017}
{Samus'}, N.~N., {Kazarovets}, E.~V., {Durlevich}, O.~V., {Kireeva}, N.~N., \& {Pastukhova}, E.~N. 2017, Astronomy Reports, 61, 80, \dodoi{10.1134/S1063772917010085}

\bibitem[{{Scicluna} {et~al.}(2015){Scicluna}, {Siebenmorgen}, {Wesson}, {Blommaert}, {Kasper}, {Voshchinnikov}, \& {Wolf}}]{Scicluna2015}
{Scicluna}, P., {Siebenmorgen}, R., {Wesson}, R., {et~al.} 2015, \aap, 584, L10, \dodoi{10.1051/0004-6361/201527563}

\bibitem[{{Shakura} \& {Sunyaev}(1973)}]{Shakura1973}
{Shakura}, N.~I., \& {Sunyaev}, R.~A. 1973, \aap, 24, 337

\bibitem[{{Shatsky} {et~al.}(2020){Shatsky}, {Belinski}, {Dodin}, {Zheltoukhov}, {Kornilov}, {Postnov}, {Potanin}, {Safonov}, {Tatarnikov}, \& {Cherepashchuk}}]{Shatsky2020}
{Shatsky}, N., {Belinski}, A., {Dodin}, A., {et~al.} 2020, in Ground-Based Astronomy in Russia. 21st Century, ed. I.~I. {Romanyuk}, I.~A. {Yakunin}, A.~F. {Valeev}, \& D.~O. {Kudryavtsev}, 127--132, \dodoi{10.26119/978-5-6045062-0-2\_2020\_127}

\bibitem[{{Shrestha} {et~al.}(2021){Shrestha}, {Neilson}, {Hoffman}, {Ignace}, \& {Fullard}}]{Shrestha2021}
{Shrestha}, M., {Neilson}, H.~R., {Hoffman}, J.~L., {Ignace}, R., \& {Fullard}, A.~G. 2021, \mnras, 500, 4319, \dodoi{10.1093/mnras/staa3508}

\bibitem[{{Skrutskie} {et~al.}(2006){Skrutskie}, {Cutri}, {Stiening}, {Weinberg}, {Schneider}, {Carpenter}, {Beichman}, {Capps}, {Chester}, {Elias}, {Huchra}, {Liebert}, {Lonsdale}, {Monet}, {Price}, {Seitzer}, {Jarrett}, {Kirkpatrick}, {Gizis}, {Howard}, {Evans}, {Fowler}, {Fullmer}, {Hurt}, {Light}, {Kopan}, {Marsh}, {McCallon}, {Tam}, {Van Dyk}, \& {Wheelock}}]{2mass}
{Skrutskie}, M.~F., {Cutri}, R.~M., {Stiening}, R., {et~al.} 2006, \aj, 131, 1163, \dodoi{10.1086/498708}

\bibitem[{{Sloan} {et~al.}(2003){Sloan}, {Kraemer}, {Price}, \& {Shipman}}]{isohpdp}
{Sloan}, G.~C., {Kraemer}, K.~E., {Price}, S.~D., \& {Shipman}, R.~F. 2003, \apjs, 147, 379, \dodoi{10.1086/375443}

\bibitem[{{Suh}(2000)}]{Suh2000}
{Suh}, K.-W. 2000, \mnras, 315, 740, \dodoi{10.1046/j.1365-8711.2000.03482.x}

\bibitem[{{Tatarnikov} {et~al.}(2024){Tatarnikov}, {Zheltoukhov}, {Shenavrin}, {Sergeenkova}, \& {Vakhonin}}]{Tatarnikov2024}
{Tatarnikov}, A.~M., {Zheltoukhov}, S.~G., {Shenavrin}, V.~I., {Sergeenkova}, I.~V., \& {Vakhonin}, A.~A. 2024, Astronomy Letters, 50, 53, \dodoi{10.1134/S1063773724600176}

\bibitem[{{Treffers} \& {Cohen}(1974)}]{Treffers1974}
{Treffers}, R., \& {Cohen}, M. 1974, \apj, 188, 545, \dodoi{10.1086/152746}

\bibitem[{{van Belle} {et~al.}(1997{\natexlab{a}}){van Belle}, {Dyck}, {Thompson}, {Benson}, \& {Kannappan}}]{vanBelle1997}
{van Belle}, G.~T., {Dyck}, H.~M., {Thompson}, R.~R., {Benson}, J.~A., \& {Kannappan}, S.~J. 1997{\natexlab{a}}, \aj, 114, 2150, \dodoi{10.1086/118635}

\bibitem[{{van Belle} {et~al.}(1997{\natexlab{b}}){van Belle}, {Dyck}, {Thompson}, {Benson}, \& {Kannappan}}]{angsize}
---. 1997{\natexlab{b}}, \aj, 114, 2150, \dodoi{10.1086/118635}

\bibitem[{{Van de Sande} {et~al.}(2024){Van de Sande}, {Walsh}, {Danilovich}, {De Ceuster}, \& {Ceulemans}}]{VandeSande2024}
{Van de Sande}, M., {Walsh}, C., {Danilovich}, T., {De Ceuster}, F., \& {Ceulemans}, T. 2024, \mnras, 532, 734, \dodoi{10.1093/mnras/stae1553}

\bibitem[{{Whitelock} {et~al.}(2008){Whitelock}, {Feast}, \& {Van Leeuwen}}]{Whitelock2008}
{Whitelock}, P.~A., {Feast}, M.~W., \& {Van Leeuwen}, F. 2008, \mnras, 386, 313, \dodoi{10.1111/j.1365-2966.2008.13032.x}

\bibitem[{{Whitney} \& {Hartmann}(1992)}]{Whitney1992}
{Whitney}, B.~A., \& {Hartmann}, L. 1992, \apj, 395, 529, \dodoi{10.1086/171673}

\bibitem[{{Wiegert} {et~al.}(2020){Wiegert}, {Groenewegen}, {Jorissen}, {Decin}, \& {Danilovich}}]{Wiegert2020}
{Wiegert}, J., {Groenewegen}, M.~A.~T., {Jorissen}, A., {Decin}, L., \& {Danilovich}, T. 2020, \aap, 642, A142, \dodoi{10.1051/0004-6361/202038029}

\bibitem[{Williams \& Rasmussen(2006)}]{Rasmussen2006}
Williams, C.~K., \& Rasmussen, C.~E. 2006 (MIT press Cambridge, MA)

\bibitem[{{Wolff} {et~al.}(2017){Wolff}, {Perrin}, {Stapelfeldt}, {Duch{\^e}ne}, {M{\'e}nard}, {Padgett}, {Pinte}, {Pueyo}, \& {Fischer}}]{Wolff2017}
{Wolff}, S.~G., {Perrin}, M.~D., {Stapelfeldt}, K., {et~al.} 2017, \apj, 851, 56, \dodoi{10.3847/1538-4357/aa9981}

\bibitem[{{Wright} {et~al.}(2010){Wright}, {Eisenhardt}, {Mainzer}, {Ressler}, {Cutri}, {Jarrett}, {Kirkpatrick}, {Padgett}, {McMillan}, {Skrutskie}, {Stanford}, {Cohen}, {Walker}, {Mather}, {Leisawitz}, {Gautier}, {McLean}, {Benford}, {Lonsdale}, {Blain}, {Mendez}, {Irace}, {Duval}, {Liu}, {Royer}, {Heinrichsen}, {Howard}, {Shannon}, {Kendall}, {Walsh}, {Larsen}, {Cardon}, {Schick}, {Schwalm}, {Abid}, {Fabinsky}, {Naes}, \& {Tsai}}]{WISE}
{Wright}, E.~L., {Eisenhardt}, P. R.~M., {Mainzer}, A.~K., {et~al.} 2010, \aj, 140, 1868, \dodoi{10.1088/0004-6256/140/6/1868}

\bibitem[{{Zheltoukhov} {et~al.}(2020){Zheltoukhov}, {Tatarnikov}, \& {Shatsky}}]{Zheltoukhov2020}
{Zheltoukhov}, S.~G., {Tatarnikov}, A.~M., \& {Shatsky}, N.~I. 2020, Astronomy Letters, 46, 193, \dodoi{10.1134/S106377372002005X}

\end{thebibliography}
\bibliographystyle{aasjournal}



\end{document}